\documentclass[prx,aps,superscriptaddress,footinbib,twocolumn]{revtex4-2}

\usepackage{amsthm,amsmath,amssymb}
\usepackage{mathrsfs}

\newcommand{\st}{\,:\,}

\usepackage{amsthm}

\theoremstyle{definition}

\theoremstyle{remark}

\usepackage{amssymb}
\usepackage{graphicx}
\usepackage{epstopdf}
\newcommand{\figpath}{./figures}

\usepackage{algorithm}
\usepackage[noend]{algpseudocode}

\usepackage{natbib}

\usepackage[colorlinks=true,citecolor=blue,linkcolor=magenta]{hyperref}

\usepackage{xcolor}

\usepackage[normalem]{ulem}

\usepackage{comment}
\usepackage{diagbox}
\usepackage{multirow}

\usepackage{pifont}
\newcommand{\cmark}{\ding{51}}
\newcommand{\xmark}{\ding{55}}
\newcommand{\omark}{\ding{109}}

\begin{document}

\title{Simple, Efficient, and Generic Post-Selection Decoding for qLDPC Codes}

\author{Haipeng Xie}
\affiliation{Graduate School of China Academy of Engineering Physics, Beijing 100193, China}

\author{Nobuyuki Yoshioka}
\affiliation{\mbox{International Center for Elementary Particle Physics, University of Tokyo, 7-3-1 Hongo, Bunkyo-ku, Tokyo 113-0033, Japan}}

\author{Kento Tsubouchi}
\affiliation{\mbox{Department of Applied Physics, University of Tokyo, 7-3-1 Hongo, Bunkyo-ku, Tokyo 113-8656, Japan}}

\author{Ying Li}
\email{yli@gscaep.ac.cn}
\affiliation{Graduate School of China Academy of Engineering Physics, Beijing 100193, China}

\begin{abstract}
Quantum error correction is indispensable for scalable quantum computation. Although encoding logical qubits substantially enhances noise resilience, achieving logical error rates low enough for practical algorithms remains challenging on existing hardware. Here we introduce argument reweighting, a simple and broadly applicable post-selection decoding strategy that boosts the performance of maximum-likelihood-type decoders, including minimum-weight perfect matching and belief-propagation families. The method suppresses logical errors by performing additional decoding rounds under reweighted error models, enabling acceptance of high-confidence syndrome outcomes. Circuit-level simulations across multiple decoders and qLDPC codes show that argument reweighting substantially suppresses logical errors, requiring a rejection rate of only $1.44\times10^{-5}$ to reduce the logical error rate by almost two orders of magnitude for the $[[144,12,12]]$ bivariate bicycle code. These results establish argument reweighting as a practical and resource-efficient approach for enhancing quantum fault tolerance.
\end{abstract}

\maketitle

{\it Introduction.---}Quantum error correction is essential for enabling many promising applications of quantum computation, such as large-integer factorization and high-accuracy simulation of quantum many-body systems~\cite{good_calderbank_1996,surface_fowler_2012,resilient_knill_1998,how_gidney_2025,encoding_babbush_2018}. A central challenge in realizing quantum error correction is the substantial qubit overhead: increasing the code distance to achieve a target logical fidelity requires more physical qubits~\cite{faulttolerant_aharonov_1999,quantum_googlequantumaiandcollaborators_2025}. Recent efforts have focused on developing low-overhead quantum error correction techniques, particularly high-rate quantum low-density parity-check (qLDPC) codes~\cite{quantum_breuckmann_2021,faulttolerant_gottesman_2014,highthreshold_bravyi_2024}. Alongside code design, improving decoding performance is equally important, as a more effective decoder can suppress logical errors without additional qubit resources~\cite{high_wootton_2012a,learning_bausch_2024,improved_muller_2025,qecgpt_cao_2023}.

\begin{figure}[t]
\centering
\includegraphics[width=1\linewidth]{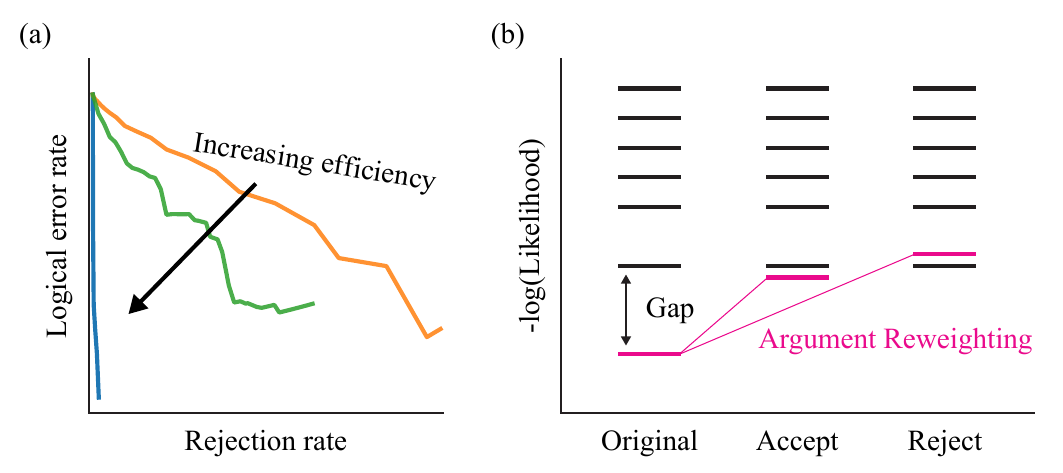}
\caption{
(a) Schematic illustration of post-selection decoding efficiency. A scheme is more efficient if it achieves the same logical-error suppression at a lower rejection rate, or stronger suppression at the same rejection rate. Thus, schemes closer to the lower-left corner are more efficient.
(b) Schematic illustration of argument reweighting (shown with the physical-error criterion). The correction corresponds to the most likely error configuration among all those compatible with the measured syndrome, which are represented by horizontal bars. The first decoding round uses the original error model and outputs the candidate correction (highlighted in magenta). In the second round, decoding is performed under a modified error model which penalizes the candidate. The circuit run is accepted if and only if the second-round output agrees with the first; otherwise, it is rejected.
}
\label{fig:scheme}
\end{figure}

Incorporating post-selection into the decoding process offers a practical means to further reduce logical errors. By using measured error syndromes and derived information, one can accept or reject a circuit run, to achieve lower logical error rates in exchange with fewer number of valid shots. In this suppression-rejection tradeoff, lower rejection for comparable logical-error suppression, or stronger suppression at fixed rejection, indicates higher post-selection efficiency; see Fig.~\ref{fig:scheme}(a). Simple and broadly applicable acceptance rules, such as rejecting runs based on detector density or correction weight, can significantly suppress logical errors, but typically incur relatively high rejection rates~\cite{realization_zhao_2022,repeated_andersen_2020,magic_rosenfeld_2025a,efficient_lee_2025}. This lead to exploration of more sophisticated rules, including boundary-edge modification for minimum-weight perfect matching (MWPM) decoders, referred to as complementary gap~\cite{efficient_hutter_2014,faulttolerant_bombin_2024,yoked_gidney_2025,mitigating_smith_2024}, and error-sparsity estimation via cluster analysis for clustering-based decoders, referred to as error-cluster statistics~\cite{efficient_lee_2025}. Although these approaches improve the post-selection efficiency, a strategy that combines a favorable suppression-rejection tradeoff with broad compatibility across different decoders and codes has yet to be established.

In this work, we introduce \emph{argument reweighting}, a simple and generic post-selection decoding scheme. It turns a decoder into a reliability test: a decoding outcome is accepted only if it remains stable when the assumed error model is deliberately perturbed; see Fig.~\ref{fig:scheme}(b). The generality stems from its compatibility with any maximum-likelihood-type decoder---those that determine corrections by exactly or approximately maximizing a likelihood (or related objective) derived from the error model. This class encompasses most widely used decoders, including MWPM and belief propagation (BP)~\cite{topological_dennis_2002,sparse_higgott_2025,1315899,Panteleev2021degeneratequantum,decoding_roffe_2020}; see Table~\ref{tab:Applicability}. Furthermore, argument reweighting is universally applicable across diverse quantum codes. The scheme is also remarkably simple, requiring only a multi-round decoding that is designed to detect the robustness of decoding outcome under reweighting of the error model, without additional qubits, circuit modifications, or code-specific redesign.

Numerical simulations using circuit-level noise models demonstrate that argument reweighting efficiently suppresses logical errors. We apply the method to a range of decoders, including MWPM and three variants of BP, and test it on rotated surface codes~\cite{surface_horsman_2012} as well as instances from the bivariate bicycle (BB) code family~\cite{highthreshold_bravyi_2024}. We observe a clear trend that the method becomes increasingly efficient with growing code distance: at larger distances, the logical error rate decreases more rapidly as the rejection rate increases.
For the largest code studied, argument reweighting reduces the logical error rate by more than a factor of $60$ while discarding only about one run in $70{,}000$.
We further compare our method with existing post-selection strategies for qLDPC codes, including detector density, correction weight, and error-cluster statistics, and find that argument reweighting outperforms them, achieving up to a reduction in logical error rate by two orders of magnitude at substantially lower rejection rates. Moreover, we show that the scheme operates seamlessly with sliding-window decoding frameworks~\cite{topological_dennis_2002,scalable_tan_2023,parallel_skoric_2023,lowlatency_gong_2024}, underscoring its direct relevance to scalable fault-tolerant quantum computation.
Together, its effectiveness at low rejection rates, simplicity, and broad compatibility make argument reweighting a practical tool for applications such as magic-state cultivation and distillation~\cite{magic_gidney_2024,hiranoEfficientMagicState2025a,ghhp-cytl,magic_litinski_2019a,distilling_xu_2026}, as well as for near-term quantum error correction experiments, where excessive rejection can otherwise become a major obstacle to scaling and implementation~\cite{superconducting_wang_2026,logical_bluvstein_2024,he2025performance,faulttolerant_wang_2024,lacroix2025scaling,rodriguez2025experimental,hong2024entangling}.

\begin{table}[h]
\centering
\begin{tabular}{ccccc}
\hline
Decoder & ~App.~ & Decoder & ~App.~ \\
\hline
MWPM~\cite{topological_dennis_2002,sparse_higgott_2025} & \cmark & BP-OSD~\cite{Panteleev2021degeneratequantum,decoding_roffe_2020} & \cmark \\
Union Find~\cite{almostlinear_delfosse_2021,9682738} & \cmark & BP-LSD~\cite{localized_hillmann_2025} & \cmark \\
BP~\cite{1315899} & \omark & BP-AC~\cite{ambiguity_wolanski_2025} & \omark \\
Mem-BP~\cite{improved_chen_2025} & \omark & Tesseract~\cite{tesseract_beni_2025} & \omark \\
BP-GD~\cite{lowlatency_gong_2024,10619083} & \omark & DTDs~\cite{decisiontree_ott_2025} & \omark \\
Relay-BP~\cite{improved_muller_2025}  & \cmark & Neural Network~\cite{learning_bausch_2024} & \xmark \\
\hline
\end{tabular}
\caption{
Applicability of argument reweighting across various quantum error correction decoders. Our scheme is supported for decoders labeled with a check mark (benchmarked in this work) or a circle (supported in principle but not yet benchmarked). Decoders marked with a cross are incompatible. Although Union Find is not a strict maximum-likelihood decoder, we demonstrate the efficacy of our scheme in this case in Appendix~\ref{app:UF}. All decoders pending benchmarking fall within the maximum-likelihood category, including five BP-based variants and the Tesseract decoder.
}
\label{tab:Applicability}
\end{table}

{\it Argument reweighting.---}We first introduce our method in the context of maximum-likelihood decoding and then apply it to two representative practical decoders: MWPM and BP-type (such as BP-OSD and Relay-BP). 

{\bf Maximum likelihood decoding.} Let $e$ denote a physical Pauli error. A stabilizer quantum error correction code defines a mapping from $e$ to the measured syndrome in parity-check measurements, denoted by $S(e)$, and a mapping from $e$ to the corresponding logical error, denoted by $L(e)$. 

In maximum-likelihood decoding~\cite{exact_cao_2025}, the correction is determined from a physical error model $M$, which specifies the probability of each error $e$, denoted $\mathrm{Pr}(e;M)$. Given an observed syndrome $s$, the decoder outputs the correction
\begin{eqnarray}
c \in \mathrm{arg\,max}_{e \,:\, S(e) = s} \mathrm{Pr}(e;M),
\end{eqnarray}
meaning that $c$ is chosen from arguments of the maximum likelihood.

In argument reweighting, a second round of decoding is performed using a modified error model $M'$, in which the likelihood of the first-round correction $c$ is suppressed in one of two ways [see Fig.~\ref{fig:scheme}(b)]:
\begin{eqnarray}
\textit{Gap test: } &&\mathrm{Pr}(c;M') = e^{-b} \, \mathrm{Pr}(c;M), \label{eq:GT} \\
\textit{Ratio test: } &&\mathrm{Pr}(c;M') = [\, \mathrm{Pr}(c;M)]^b, \label{eq:RT}
\end{eqnarray}
where $b$ is a tunable parameter that controls the strength of the suppression, with $b>0$ for gap test and $b>1$ for ratio test. The definitions in Eqs.~\eqref{eq:GT} and \eqref{eq:RT} specify only how the likelihood assigned to the first-round correction $c$ is suppressed. To fully define the modified model $M'$, one must also specify how $M'$ treats other errors. This leads to two variants of argument reweighting. In \emph{ideal reweighting}, the probabilities of all other errors sharing the syndrome $s$ are kept unchanged. In \emph{practical reweighting}, this condition is relaxed: the weights are modified to suppress $c$, possibly also changing those of other errors with the same syndrome, as in the practical protocol defined later; see Appendix~\ref{app:IR_PR} for more discussions. Under the modified model, maximum-likelihood decoding produces a second correction
\begin{eqnarray}
c' \in \mathrm{arg\,max}_{e \,:\, S(e) = s} \mathrm{Pr}(e;M').
\end{eqnarray}
The circuit run is accepted according to the criterion:
\begin{itemize}
\item {\it Physical-error criterion (PEC)}---Accept iff $c' = c$.
\end{itemize}

We now explain why our approach reduces the logical error rate. The total logical error rate can be expressed as the average of the logical error rates conditioned on the observed syndrome $s$. The goal of post-selection is to reject those circuit runs whose syndrome $s$ is associated with a high conditional logical error rate, $p_L(s)$. Importantly, $p_L(s)$ is upper-bounded by the probability that the true error configuration differs from the decoder's output $c$. Consequently, the conditional logical error rate satisfies: 
\begin{eqnarray}
    p_L(s) \leq (m-1)e^{-\Delta},
\end{eqnarray}
where $m$ is the total number of error configurations consistent with syndrome $s$, and $\Delta$ represents the log-likelihood gap between the most probable error configuration and the next most probable one [see Fig.~\ref{fig:scheme}(b) and Appendix~\ref{app:bounds} for a proof]. Therefore, a large log-likelihood gap implies high confidence that the conditional logical error rate is small. Gap test, together with PEC, is designed to detect such gaps and thereby identify high-confidence syndromes for acceptance. We note that complementary gap decoding~\cite{efficient_hutter_2014,faulttolerant_bombin_2024,yoked_gidney_2025,mitigating_smith_2024}, a post-selection scheme for surface codes, operates on a similar principle: it probes the likelihood gap by modifying boundary edges of the decoding graph.

In our framework, ratio test serves an analogous purpose to gap test. Numerical results show that both mechanisms are effective, with ratio test slightly outperforms gap test: for the $[[72,12,6]]$ BB code decoded via BP-OSD, the ratio test achieves a tenfold suppression of the logical error rate with a rejection rate approximately $18\%$ lower than that of the gap test; see Appendix~\ref{app:GT} for details.

In addition to the physical-error criterion, we find that certain variants can further improve performance. Since all corrections within the same logical equivalence class produce the same logical effect, we may relax the acceptance condition to
\begin{itemize}
\item {\it Two-round logical-error criterion (2R-LEC)}---Accept iff $L(c') = L(c)$.
\end{itemize}
We may also strengthen the confidence of acceptance by performing a third round of decoding, using a reweighted error model $M''$ that suppresses the likelihoods of both $c$ and $c'$. Let $c''$ be the output of the third-round decoding under $M''$. This leads to an enhanced rule:
\begin{itemize}
\item {\it Three-round logical-error criterion (3R-LEC)}---Accept iff $L(c'') = L(c') = L(c)$.
\end{itemize}
Note that one can further increase the number of reweighting-decoding rounds.

\begin{figure*}[t]
\centering
\includegraphics[width=1\linewidth]{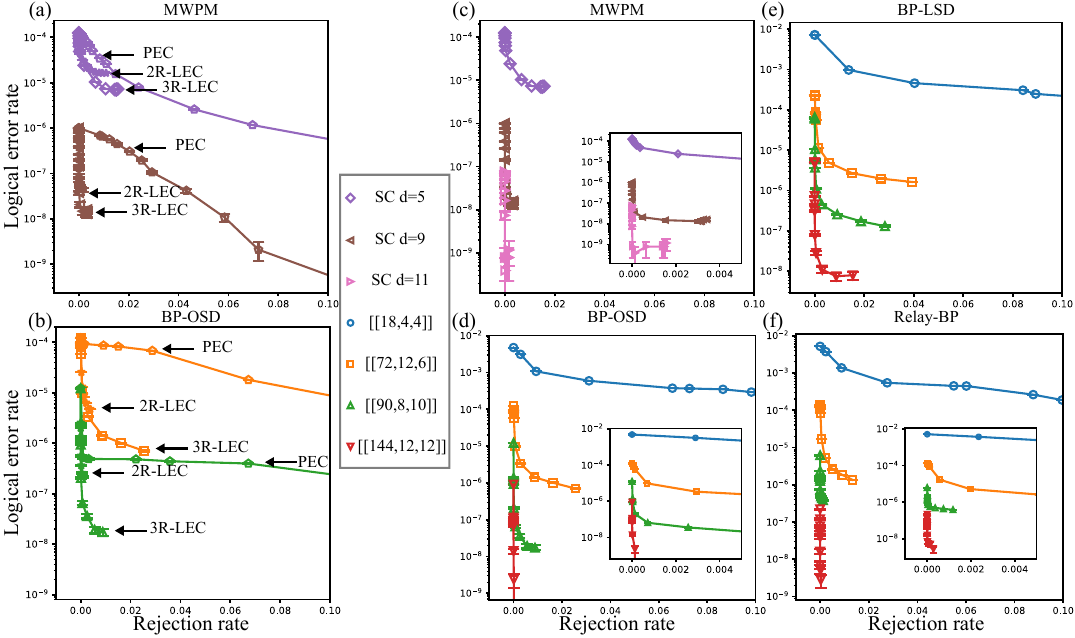}
\caption{
Total logical error rate after $d$ rounds of parity-check measurements, where $d$ denotes the code distance. We simulated rotated surface codes (SC) with $d=5, 9,$ and $11$~\cite{surface_horsman_2012}, as well as four instances of BB codes: $[[18,4,4]]$, $[[72,12,6]]$, $[[90,8,10]]$, and $[[144,12,12]]$~\cite{highthreshold_bravyi_2024,coprime_wang_2025}. Results are obtained using the ratio test variant defined in Eq.~\eqref{eq:reweight}. Further details regarding the simulation are provided in Appendix~\ref{app:details}.
}
\label{fig:Logical_error}
\end{figure*}

{\bf MWPM.} The decoding is performed on a weighted graph $(V, E, w)$, where $V$ is the set of nodes representing parity checks, $E$ is the set of edges corresponding to elementary errors, and $w \st E \to \mathbb{R}$ assigns a weight to each edge. In the error model, an elementary error $q \in E$ occurs with probability $p(q)$. The edge weights are defined as $w(q) = \ln[1/p(q)-1]$, effectively mapping error probabilities to additive weights~\cite{topological_dennis_2002}. A physical error configuration $e$ is represented by a subset of edges $e \subseteq E$, and an observed syndrome $s$ corresponds to a subset of nodes $s \subseteq V$. MWPM aims to find a set of paths that pair up all nodes in $s$ while minimizing the total weight, which is equivalent to maximizing the likelihood. The union of the edges on these paths constitutes the correction $c \subseteq E$.

As a practical realization of the ratio-test argument reweighting in Eq.~\eqref{eq:RT}, we update the elementary error probabilities from $p(q)$ to $p'(q)$:
\begin{eqnarray}
p'(q) =
\begin{cases}
p(q)^b, & q \in c, 
\\[6pt]
p(q), & q \notin c,
\end{cases}
\label{eq:reweight}
\end{eqnarray}
where $b>1$ is the reweighing parameter. Note that the likelihood modification in Eq.~\eqref{eq:reweight} satisfies Eq.~\eqref{eq:RT} to a good approximation; an exact variant and the gap-test version are given in Appendices~\ref{app:ERT} and~\ref{app:GT}. Because this elementary-error update also modifies the likelihoods of other candidate errors that overlap with $c$, the protocol falls into the practical, rather than ideal, reweighting class. Unless otherwise stated, in what follows ``argument reweighting'' refers to this practical elementary-error realization. When a second reweighting round is performed, the updated probabilities $p''$ are derived analogously by modifying $p'$ based on the second-round output $c'$.

{\bf BP.} The decoding is performed on a Tanner (bipartite) graph $(C, B, E)$, where $C$ is the set of nodes corresponding to parity checks, $B$ is the set of nodes representing elementary errors, and $E \subseteq C \times B$ specifies which elementary errors are detected by which checks. As in MWPM decoding, an elementary error $q \in B$ occurs with probability $p(q)$; a physical error configuration is a subset $e \subseteq B$; and a measured syndrome is a subset $s \subseteq C$. Given $s$, BP iteratively exchanges messages between $C$ and $B$, updating marginal likelihoods until convergence (or until a maximum number of iterations is reached). A final decision is then made, yielding a correction $c \subseteq B$.

In argument reweighting, we modify the error model in the same manner as in MWPM, using Eq.~\eqref{eq:reweight}, with the only difference that the correction $c$ is now a subset of $B$. Pseudocodes for MWPM and BP decoding corresponding to the three acceptance criteria are provided in Appendix~\ref{app:codes}.

{\it Numerical performance and benchmarks}.---We evaluate the performance of PEC and LEC on both rotated surface codes and BB codes using their respective mainstream decoders. Simulations are performed under a standard circuit-level noise model with physical error rate $p = 0.1\%$; see Appendix~\ref{app:details} for details. As shown in Figs.~\ref{fig:Logical_error}(a) and (b), under PEC, the logical error rates of both rotated surface codes and BB codes decrease steadily as the rejection rate increases, for both MWPM and BP-OSD decoders. As a more permissive strategy, LEC substantially improves post-selection efficiency in the low-rejection-rate regime, which is the most relevant in practice. However, under LEC, the reduction in the logical error rate, following an initial sharp drop, eventually slows down and approaches saturation: as observed in Fig.~\ref{fig:Logical_error}(c), beyond a certain rejection rate, additional post-selection yields no further improvement. This saturation threshold can be pushed by increasing the number of reweighting-decoding rounds, and employing additional rounds can yield further reductions in the logical error rate. In addition to the 2R-LEC and 3R-LEC results shown in Fig.~\ref{fig:Logical_error}, we provide data for extra rounds and a detailed explanation of the saturation phenomenon in Appendix~\ref{app:explanation}.

Figs.~\ref{fig:Logical_error}(c)-(f) illustrate the performance of 3R-LEC across a range of decoders, including MWPM, BP-OSD, BP-LSD, and Relay-BP, evaluated on additional code examples. We observe that 3R-LEC becomes increasingly efficient with growing code distance: at larger distances, the logical error rate decreases more rapidly as the rejection rate increases; see Appendxi~\ref{app:LESF} for plots of rescaled logical error rates, which illustrate the post-selection efficiency more transparently. Furthermore, 3R-LEC consistently achieves a suppression of logical errors by one order of magnitude across all tested decoders and code instances, and reaches two orders of magnitude for codes with sufficiently large distances. For the BP-OSD (Relay-BP) decoder applied to the $[[144,12,12]]$ BB code, argument reweighting reduces the logical error rate from $9.08\times10^{-7}$ $(2.20\times10^{-7})$ to $1.41\times10^{-8}$ $(1.60\times10^{-8})$ at a rejection rate of only $1.44\times10^{-5}$ $(1.83\times10^{-6})$ after $12$ rounds of parity-check measurements.

{\it Comparisons with other post-selection strategies.---}
A key advantage of our post-selection strategy over complementary gap~\cite{efficient_hutter_2014,faulttolerant_bombin_2024,yoked_gidney_2025,mitigating_smith_2024} and error-cluster statistics~\cite{efficient_lee_2025} is its broad applicability across decoders, rather than being limited to a specific family.

\begin{figure}[t]
\centering
\includegraphics[width=1\linewidth]{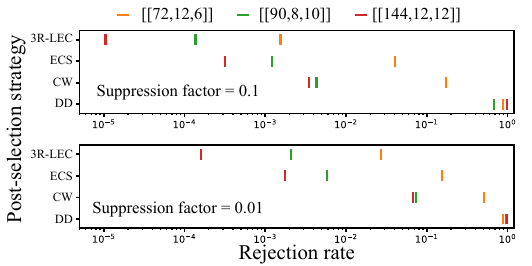}
\caption{
Efficiency of various post-selection strategies applied to the BP-LSD decoder. The performance of four strategies---3R-LEC, error-cluster statistics (ECS), correction weight (CW), and detector density (DD)~\cite{efficient_lee_2025}---is compared across three BB code instances. For each code, we estimate the rejection rates (per $d$ parity-check cycles) required to reduce the logical error rate by one and two orders of magnitude. We find that our method (3R-LEC) consistently requires the lowest rejection rate among all post-selection strategies considered. We note that the rejection rates for 3R-LEC are reported as conservative upper bounds, whereas those for other strategies are calculated precisely (subject to statistical error); see Ref.~\cite{note1} and Appendix~\ref{app:details} for further details.
}
\label{fig:Compare}
\end{figure}

In terms of efficiency, when applied to BB codes, 3R-LEC significantly outperforms error-cluster statistics, achieving a suppression of logical errors by one to two orders of magnitude at substantially lower rejection rates (Fig.~\ref{fig:Compare}). We note that this comparison is performed using the BP-LSD decoder, while our 3R-LEC scheme is also compatible with other decoders for qLDPC codes. Complementary gap is currently inapplicable to general qLDPC codes. For rotated surface codes, complementary gap achieves higher efficiency, as it estimates the likelihood gap more accurately via decoding-graph modifications. Specifically, for the $d=9$ rotated surface code decoded via MWPM, the complementary gap method achieves a tenfold suppression of the logical error rate with a rejection rate approximately $41\%$ lower than that of 3R-LEC (see Appendix~\ref{app:CG} for more results).

Regarding runtime, our method requires only a small, fixed number of decoding rounds (at most three in practice, as shown in our numerical results), incurring a moderate computational overhead. We note that reweighting affects the BP convergence rate, which is discussed in Appendix~\ref{app:convergence}. In comparison, while error-cluster statistics operates with a single decoding round, the runtime of complementary gap scales exponentially with the number of logical qubits $k$. Specifically, complementary gap requires $2^k$ rounds to evaluate all logical equivalence classes, rendering it impractical for multi-logical-qubit scenarios. Such cases include high-rate code blocks encoding multiple logical qubits~\cite{quantum_breuckmann_2021,faulttolerant_gottesman_2014,highthreshold_bravyi_2024}, as well as the collective decoding of transversal controlled-NOT gate sequences or lattice surgery operations involving multiple surface-code blocks~\cite{lowoverhead_zhou_2025,game_litinski_2019}. In contrast, our strategy maintains a small number of decoding rounds, remaining computationally tractable even for the 12-logical-qubit cases examined in our numerical results.

{\it Remark on applicability.---}Argument reweighting is a framework with broad utility in fault-tolerant quantum computing. It is compatible with sliding-window decoding~\cite{topological_dennis_2002,scalable_tan_2023,parallel_skoric_2023,lowlatency_gong_2024}, well-suited for large-scale implementations, and can seamlessly incorporate soft-syndrome information~\cite{improved_pattison_2021,soft_raveendran_2022}. Further discussions 
and supporting numerical results are provided in Appendices~\ref{app:SWD},~\ref{app:scalability} and~\ref{app:soft-info}.

{\it Conclusions.---}In this work, we introduced argument reweighting, a post-selection decoding strategy applicable to a broad class of decoders and quantum error correction codes. We benchmarked its performance using circuit-level simulations across multiple decoders and codes, and demonstrated compatibility with sliding-window decoding. Among the three rejection criteria we propose, 3R-LEC offers the best trade-off between logical error suppression and rejection rate in the low-rejection regime. Our numerical results show that the method typically yields one to three orders of magnitude reduction in logical error rate (see Fig.~\ref{fig:Logical_error}), while requiring significantly lower rejection rates than existing post-selection approaches that are scalably applicable to qLDPC codes encoding multiple logical qubits (see Fig.~\ref{fig:Compare}). These results establish argument reweighting as a practical and efficient tool for enhancing logical reliability, particularly in early fault-tolerant quantum computing architectures where resource efficiency is critical.

\begin{acknowledgments}
This work is supported by the National Natural Science Foundation of China (Grant Nos. 12225507, 12088101) and NSAF (Grant No. U1930403).
N.Y. is supported by JST Grant Number JPMJPF2221, JST CREST Grant Number JPMJCR23I4, IBM Quantum, JST ASPIRE Grant Number JPMJAP2316, JST ERATO Grant Number JPMJER2302, JST [Moonshot R\&D] [Grant Number JPMJMS256J], and Institute of AI and Beyond of the University of Tokyo. The source codes for the numerical simulation are available at~\cite{Code_Xie}.
\end{acknowledgments}

\appendix

\begin{widetext}

\section{Bounds on the conditional logical error rate}
\label{app:bounds}

The probability of observing syndrome $s$ is
\begin{equation}
\mathrm{Pr}(s) = \sum_{e \st S(e) = s} \mathrm{Pr}(e; M).
\end{equation}

For the raw maximum-likelihood decoding, the logical error probability conditioned on $s$ is
\begin{equation}
p_L(s) = 1 - \frac{\sum_{e \st S(e) = s,\, L(e) = L(c)} \mathrm{Pr}(e; M)}{\mathrm{Pr}(s)},
\end{equation}
where $c$ is the correction produced by the decoder with the input syndrome $s$. This probability is upper bounded by
\begin{equation}
p_L(s) \leq 1 - \frac{\mathrm{Pr}(c; M)}{\mathrm{Pr}(s)}.
\end{equation}
The total logical error probability without post-selection is
\begin{equation}
p_{L, \mathrm{tot}} = \sum_s \mathrm{Pr}(s) \, p_L(s).
\end{equation}

Let $e_\mathrm{next}$ denote the next most probable error, i.e.,
\begin{equation}
e_\mathrm{next} \in \mathrm{arg\,max}_{e \,:\, S(e) = s,\, e \neq c} \mathrm{Pr}(e; M),
\end{equation}
and define the likelihood gap $\Delta$ by
\begin{equation}
\mathrm{Pr}(e_\mathrm{next}; M) = e^{-\Delta} \, \mathrm{Pr}(c; M).
\end{equation}

Let $m = \vert\{ e \st S(e) = s \}\vert$ be the total number of errors with syndrome $s$. Then
\begin{equation}
\frac{\mathrm{Pr}(c; M)}{\mathrm{Pr}(s)} \geq 
\frac{\mathrm{Pr}(c; M)}{\mathrm{Pr}(c; M) + (m-1)\mathrm{Pr}(e_\mathrm{next}; M)} 
= \frac{1}{1 + (m-1)e^{-\Delta}},
\end{equation}
and consequently
\begin{equation}
p_L(s) \leq 1 - \frac{1}{1 + (m-1)e^{-\Delta}} \leq (m-1)e^{-\Delta}.
\end{equation}

\section{Pseudocodes}
\label{app:codes}

The algorithms in this section utilize the following notation:
\begin{itemize}
\item $\textsc{Decoder}(p, s) \to c$: A maximum-likelihood-type decoder that accepts an error model $p$ and a syndrome $s$ as input. Here, $p: E \to \mathbb{R}$ specifies the prior probabilities for the set of elementary errors $E$, and $c$ is the resulting physical correction.
\item Logical Map $L$: A function that maps a physical error (or correction) to its corresponding logical error class.
\item Post-Selection Strategy $R$: The specific scheme employed, where $R \in \{\text{PEC, 2R-LEC, 3R-LEC}\}$.
\item Reweighting Parameter $b$: The tunable parameter used in the modified error models to control the strength of suppression during subsequent decoding rounds.
\end{itemize}

\begin{algorithm}[H]
\caption{Post-Selection Decoding via Argument Reweighting}
\label{alg:argument_reweighting}
\begin{algorithmic}[1]
\Statex
\Function{ArgumentReweighting}{\textsc{Decoder}, $p$, $s$, $L$, $R$, $b$}
    \State $c \gets \textsc{Decoder}(p, s)$ \Comment{Obtain first-round correction based on priors $p$ and syndrome $s$}
    
    \If{$c = \emptyset$} \Comment{If no error is detected, the run is accepted by default}
        \State \Return Accept, $c$
    \EndIf
    
    \State $\text{Flag} \gets \Call{PostSelection}{\textsc{Decoder}, p, s, L, R, b, c}$
    
    \If{$\text{Flag} = 1$}
        \State \Return Accept, $c$
    \Else
        \State \Return Reject
    \EndIf
\EndFunction
\end{algorithmic}
\end{algorithm}

\begin{algorithm}[H]
\caption{Post-Selection}
\label{alg:post_selection}
\begin{algorithmic}[1]
\Statex
\Function{PostSelection}{\textsc{Decoder}, $p$, $s$, $L$, $R$, $b$, $c$}
    \State $\text{Flag} \gets 0$
    \State $p' \gets \Call{Reweighting}{p, b, c}$ \Comment{Generate first modified error model}
    \State $c' \gets \textsc{Decoder}(p', s)$ \Comment{Second round of decoding}
    
    \If{$R = \text{PEC}$} \Comment{Physical-Error Criterion}
        \If{$c' = c$}
            \State $\text{Flag} \gets 1$
        \EndIf
        
    \ElsIf{$R = \text{2R-LEC}$} \Comment{Two-Round Logical-Error Criterion}
        \If{$L(c') = L(c)$}
            \State $\text{Flag} \gets 1$
        \EndIf
        
    \ElsIf{$R = \text{3R-LEC}$} \Comment{Three-Round Logical-Error Criterion}
        \If{$L(c') = L(c)$}
            \State $p'' \gets \Call{Reweighting}{p', b, c'}$ \Comment{Generate second modified error model}
            \State $c'' \gets \textsc{Decoder}(p'', s)$ \Comment{Third round of decoding}
            \If{$L(c'') = L(c)$}
                \State $\text{Flag} \gets 1$
            \EndIf
        \EndIf
    \EndIf
    
    \State \Return $\text{Flag}$
\EndFunction
\end{algorithmic}
\end{algorithm}

\begin{algorithm}[H]
\caption{Reweighting---Eq.~\eqref{eq:reweight} in main text}
\label{alg:reweighting}
\begin{algorithmic}[1]
\Statex
\Function{Reweighting}{$p, b, c$}
    \State $p' \gets p$ \Comment{Initialize the modified error model}
    \For{each elementary error $q \in c$}
        \State $p'(q) \gets p(q)^b$ \Comment{Suppress the likelihood of the previous correction}
    \EndFor
    \State \Return $p'$
\EndFunction
\end{algorithmic}
\end{algorithm}

\section{Details of numerical simulations}
\label{app:details}

\subsection{Circuits and noise model}
Simulations are performed using rotated surface-code circuits based on the \textit{Stim} library~\cite{gidney2021stim} and BB code circuits from Ref.~\cite{lowlatency_gong_2024}. We assume a circuit-level noise model with a physical error rate of $p=0.001$, characterized by the following:
\begin{itemize}
    \item Single-qubit Clifford gates are subject to depolarizing noise with probability $p$, where each Pauli error $X$, $Y$, or $Z$ occurs with probability $p/3$.
    \item Two-qubit Clifford gates are subject to two-qubit depolarizing noise with probability $p$; each of the 15 non-trivial two-qubit Pauli errors occurs with probability $p/15$.
    \item Idling qubits undergo depolarizing noise with probability $p$ during each time step.
    \item State preparation and resets in the $Z$ ($X$) basis are followed by a bit-flip $X$ (phase-flip $Z$) error with probability $p$.
    \item Measurement outcomes are flipped with probability $p$.
\end{itemize}
All syndrome sampling and circuit simulations are conducted using the \textit{Stim} library.

\subsection{Decoder implementation}

We utilize several open-source libraries for our decoder implementations: MWPM is implemented via \textit{PyMatching}~\cite{sparse_higgott_2025}, BP-OSD and BP-LSD are sourced from the \textit{ldpc} library~\cite{decoding_roffe_2020}, and the Relay-BP decoder is implemented using the \textit{relay} library~\cite{improved_muller_2025}. We employ optimized parameter configurations for each decoder to ensure a high-performance comparison.

For MWPM, we utilize the default settings~\cite{sparse_higgott_2025}.

For BP-OSD, we take the following parameters~\cite{lowlatency_gong_2024}:
\begin{itemize}
    \item \texttt{max\_iter}: 200
    \item \texttt{bp\_method}: ``minimum\_sum''
    \item \texttt{ms\_scaling\_factor}: 1.0
    \item \texttt{osd\_method}: ``OSD\_CS''
    \item \texttt{osd\_order}: 10
\end{itemize}
For BP-LSD, we take the following parameters~\cite{localized_hillmann_2025,efficient_lee_2025}:
\begin{itemize}
    \item \texttt{max\_iter}: 30
    \item \texttt{bp\_method}: ``minimum\_sum''
    \item \texttt{ms\_scaling\_factor}: 1.0
    \item \texttt{lsd\_method}: ``LSD\_0''
    \item \texttt{lsd\_order}: 0
\end{itemize}
For Relay-BP, we take the following parameters~\cite{improved_muller_2025,fpgatailored_maurya_2025,realtime_maurer_2025}:
\begin{itemize}
    \item \texttt{gamma0}: 0.125
    \item \texttt{pre\_iter}: 80
    \item \texttt{num\_sets}: 300
    \item \texttt{set\_max\_iter}: 60
    \item \texttt{gamma\_dist\_interval}: (-0.24, 0.66)
    \item \texttt{stop\_nconv}: 1
\end{itemize}

\subsection{Post-selection parameters and sample sizes}
\label{sec:parameters}

By varying the control parameter $b = 1 + z$, we obtain post-selected logical error rates across a range of rejection rates. For the rotated surface code results shown in Fig.~\ref{fig:Logical_error} of the main text, we employ a consistent set of 17 values for $z$: $\{10^{-12}, 10^{-6}, 10^{-3}, 0.01, 0.02, 0.04, 0.06, 0.1, 0.2, 0.3, 0.4, 0.5, 0.6, 0.7, 0.8, 0.9, 1\}$. For the BB codes, 10 distinct values of $z$ are used. Specifically, we set $z \in \{10^{-8}, 0.1, 0.5, 1, 1.5, 2, 2.5, 3, 3.5, 4\}$ for the $[[18,4,4]]$ code, and $z \in \{10^{-15}, 10^{-8}, 10^{-4}, 10^{-3}, 0.01, 0.1, 0.2, 0.3, 0.4, 0.5\}$ for the remaining BB codes. For any given code, the parameter $b$ remains identical across all tested decoders.

To ensure high statistical precision, we sampled $2.56 \times 10^9$ instances for each code-decoder pair. For each value of $b$, the post-selected logical error rate and its standard deviation are estimated from the retained instances. We note that higher values of $b$ lead to increased rejection rates, thereby reducing the effective sample size of the retained data.

Due to finite sampling, the rejection rate is subject to statistical uncertainty, much like the logical error rate. Since the error bars for the rejection rate are typically negligible, they are omitted from the plots for visual clarity.

\subsection{Comparison between post-selection strategies}

\begin{figure*}
\begin{minipage}{\linewidth}
\begin{table}[H]
\begin{tabular}{ccccccc}
\hline
$\eta$ & Code & Baseline & 3R-LEC & ECS & CW & DD \\
\hline
\multirow{6}{*}{$10^{-1}$} & \multirow{2}{*}{[[72,12,6]]} & $0$ & $1.5435(8)\times10^{-3}$ & $4.0964(4)\times10^{-2}$ & $1.74331(7)\times10^{-1}$ & $8.83231(6)\times10^{-1}$ \\
\cline{3-7}
& & $2.379(3)\times10^{-4}$ & $1.145(7)\times10^{-5}$ & $2.38(1)\times10^{-5}$ & $2.31(2)\times10^{-5}$ & $0$ \\
\cline{2-7}
& \multirow{2}{*}{[[90,8,10]]} & $0$ & $1.370(2)\times10^{-4}$ & $1.2003(7)\times10^{-3}$ & $4.312(1)\times10^{-3}$& $6.78411(9)\times10^{-1}$ \\
\cline{3-7}
& & $6.49(2)\times10^{-5}$& $5.91(5)\times10^{-6}$ & $6.49(5)\times10^{-6}$ & $6.5(8)\times10^{-6}$ & $6.11(6)\times10^{-6}$\\
\cline{2-7}
& \multirow{2}{*}{[[144,12,12]]} & $0$ & $1.051(6)\times10^{-5}$ & $3.142(4)\times10^{-4}$ & $3.473(1)\times10^{-3}$ & $9.999159(2)\times10^{-1}$ \\
\cline{3-7}
& & $4.98(4)\times10^{-6}$ & $4.6(1)\times10^{-7}$ & $5.0(1)\times10^{-7}$ & $5(2)\times10^{-7}$& $0$ \\
\hline
\multirow{6}{*}{~~~$10^{-2}$~~~} & \multirow{2}{*}{[[72,12,6]]} & $0$ & $2.6838(3)\times10^{-2}$ & $1.56154(7)\times10^{-1}$ & $5.07449(10)\times10^{-1}$ & $8.83231(6)\times10^{-1}$\\
\cline{3-7}
& & $2.379(3)\times10^{-4}$ & $1.97(3)\times10^{-6}$ & $2.38(3)\times10^{-6}$ & $1.29(10)\times10^{-7}$ & $0$ \\
\cline{2-7}
& \multirow{2}{*}{[[90,8,10]]} & $0$ & $2.0949(9)\times10^{-3}$ & $5.762(1)\times10^{-3}$ & $7.4081(5)\times10^{-2}$ & $9.52405(4)\times10^{-1}$ \\
\cline{3-7}
& & $6.49(2)\times10^{-5}$ & $5.4(1)\times10^{-7}$ & $6.5(2)\times10^{-7}$ & $6.4(6)\times10^{-7}$ & $5.4(1)\times10^{-7}$\\
\cline{2-7}
& \multirow{2}{*}{~~~[[144,12,12]]~~~} & $0$ & $1.614(3)\times10^{-4}$ & $1.7750(8)\times10^{-3}$ & $6.8386(5)\times10^{-2}$ & $9.999159(2)\times10^{-1}$ \\
\cline{3-7}
& & $4.98(4)\times10^{-6}$ & $4.6(4)\times10^{-8}$& $5.0(4)\times10^{-8}$ & $5(2)\times10^{-8}$ & $0$ \\
\hline
\end{tabular}
\caption{
Numerical results for the BP-LSD decoder using the 3R-LEC, error-cluster statistics (ECS), correction weight (CW), and detector density (DD) methods. For each code, the upper (lower) row reports the corresponding rejection rate (logical error rate). These values are extracted from the same dataset used for Fig.~\ref{fig:Logical_error} in the main text. See the text for an explanation regarding the zero observed logical error rates for the DD method.
}
\label{tab:BP-LSD}
\end{table}
\end{minipage}
\end{figure*}

To evaluate the efficiency of different post-selection strategies, we define two primary metrics: (1) the rejection rate required to achieve a tenfold reduction (one order of magnitude) in the logical error rate, and (2) the rejection rate required for a hundredfold reduction (two orders of magnitude). Strategies that achieve these targets at lower rejection rates are considered more efficient.

We determine the rejection rates corresponding to tenfold ($\eta = 0.1$) and hundredfold ($\eta = 0.01$) reductions in the logical error rate as follows. First, we evaluate the baseline logical error rate $p_L$ and its standard deviation $\sigma_L$ without post-selection. For a given post-selection configuration, let $p_L'$ and $\sigma_L'$ denote the resulting logical error rate and its associated standard deviation, respectively. We define the target $\eta p_L$ as being ``achieved'' if the two rates are within one standard deviation, i.e.,
\begin{equation}
\vert \eta p_L - p_L' \vert \leq \sqrt{(\eta \sigma_L)^2 + {\sigma_L'}^2}.
\end{equation}
The target is considered ``surpassed'' if $p_L'$ is significantly lower than the target, satisfying the condition:
\begin{equation}
\eta p_L - p_L' > \sqrt{(\eta \sigma_L)^2 + {\sigma_L'}^2}.
\end{equation}

For the 3R-LEC scheme, the post-selection process is controlled by the parameter $b$. Because evaluating each value of $b$ requires a complete execution of the decoding algorithm, an exhaustive search for a configuration that exactly achieves the target is time consuming. Consequently, we terminate the parameter search once we identify a configuration that achieves or surpasses the target while remaining close to it.

The aforementioned approach provides a conservative upper bound for the rejection rate at the target logical error rate, denoted as $r_1$. To quantify the tightness of this estimate, we apply PCHIP interpolation to the BP-LSD data in Fig.~\ref{fig:Logical_error_norm} to obtain a secondary estimate, $r_2$. The relative deviation, defined as $|r_1 - r_2| / r_2$, ranges between $2\%$ and $83\%$ across the tested instances.

In contrast, for the other three strategies---error-cluster statistics (ECS), correction weight (CW), and detector density (DD)---adjusting the post-selection threshold does not require additional decoding rounds. This allows us to precisely identify a configuration that achieves the target logical error rate for ECS and CW. The case of DD is unique: here, the rejection rate is controlled by a threshold on the syndrome weight (the number of triggered detectors). In our simulations, we frequently encountered scenarios where the target logical error rate could not be met at any finite threshold, and the only remaining option is a stringent acceptance rule that rejects all instances with a non-zero syndrome weight. Such a rule results in extremely low logical error rates, yielding zero observed logical errors within our finite sample size. Therefore, for DD, we report the configuration that achieves the target rate when possible; otherwise, we report the results of the zero-weight acceptance rule. The resulting data are presented in Table~\ref{tab:BP-LSD}.

In Table~\ref{tab:BP-LSD}, the baseline, 3R-LEC, CW, and DD are evaluated based on the raw BP-LSD decoder provided by the \textit{ldpc} library, while ECS is evaluated using the \textit{ldpc-post-selection} library, which was developed in Ref.~\cite{efficient_lee_2025} to implement the BP-LSD decoder with the ECS approach. We find that a discrepancy exists between the logical error rates of the standard \textit{ldpc} library and those of the \textit{ldpc-post-selection} library, even when the latter is operated without post-selection. Specifically, the \textit{ldpc-post-selection} implementation yields logical error rates that are higher by a factor of about two across the three code instances, suggesting potential room for improvement.

\section{Logical error suppression factor}
\label{app:LESF}
\begin{figure*}[h]
\centering
\includegraphics[width=1\linewidth]{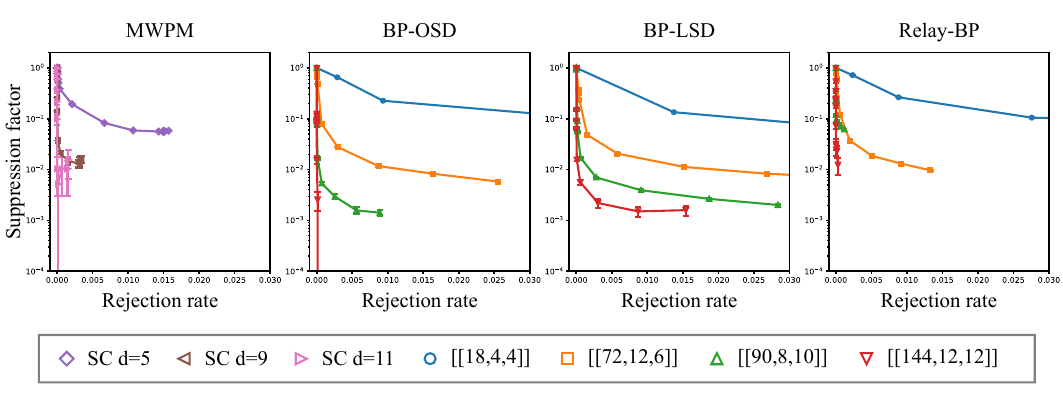}
\caption{
Logical error suppression factor. For each decoder and code, the suppression factor is defined as the logical error rate at a given rejection rate divided by the logical error rate without post-selection.
}
\label{fig:Logical_error_norm}
\end{figure*}

In Figs.~\ref{fig:Logical_error}(c)--(f) of the main text, we illustrate the relationship between the logical error rate and the rejection rate of 3R-LEC across various codes and decoders. To better evaluate the performance of 3R-LEC, Fig.~\ref{fig:Logical_error_norm} presents these results in a normalized form, where the logical error rates are expressed relative to the baseline error rates without post-selection.

\section{Post-selection in Union Find decoding}
\label{app:UF}
\begin{figure}[h]
\centering
\includegraphics[width=1\linewidth]{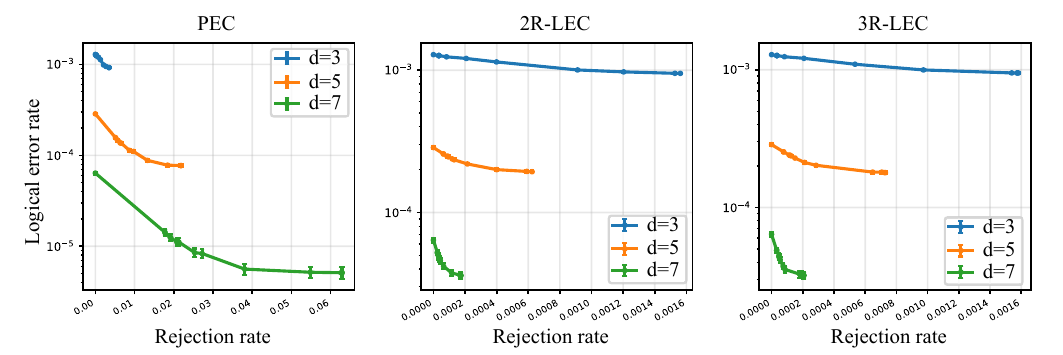}
\caption{
Logical error rates for the Union Find decoder integrated with post-selection. Each data point is estimated from $10^7$ Monte Carlo shots under circuit-level noise ($p=0.001$) with $d$ parity-check measurement rounds, where $d$ denotes the code distance of the rotated surface code. In post-selection, we take the reweighing parameter $b = 1 + z$ for ratio test, with $z \in \{10^{-12}, 10^{-6}, 10^{-3}, 0.01, 0.02, 0.04, 0.06, 0.1, 0.2, 0.3, 0.4, 0.5\}$.
}
\label{fig:UF}
\end{figure}

Given that the Union Find decoder utilizes edge weights to guide its cluster-growth process~\cite{almostlinear_delfosse_2021,9682738,improved_pattison_2021}, we anticipate its behavior under argument reweighting will be similar to that of a maximum-likelihood-type decoder.

To evaluate the applicability of argument reweighting to the Union Find decoder, we conducted benchmark simulations, the results of which are presented in Fig.~\ref{fig:UF}. We find that argument reweighting effectively reduces the logical error rate for the rotated surface code, with suppression efficiency scaling favorably with the code distance---consistent with the trends observed in the main text. However, in contrast to the maximum-likelihood-type decoders studied in the main text, 3R-LEC yields no significant improvement over 2R-LEC for the Union Find decoder.

\section{Exact ratio test}
\label{app:ERT}
The practical ratio test in Eq.~\eqref{eq:reweight} of the main text only approximately realizes the ratio test defined in Eq.~\eqref{eq:RT}. For independent elementary errors, the likelihood of the correction $c$ in the original error model is
\begin{eqnarray}
\mathrm{Pr}(c;M) =
\left[\prod_{q \in c} p(q)\right]
\left[\prod_{q \notin c} \left(1-p(q)\right)\right].
\end{eqnarray}
Under the probability update in Eq.~\eqref{eq:reweight} of the main text, the modified likelihood becomes
\begin{eqnarray}
\mathrm{Pr}(c;M') =
\left[\prod_{q \in c} p(q)^b\right]
\left[\prod_{q \notin c} \left(1-p(q)\right)\right].
\end{eqnarray}
For small error probabilities $p(q)$, the no-error factors are close to unity, so this modified likelihood approximates $[\mathrm{Pr}(c;M)]^b$, as required by Eq.~\eqref{eq:RT} of the main text.

To remove this approximation, one can instead use the update rule
\begin{eqnarray}
p'(q) =
\begin{cases}
p(q)^b, & q \in c, 
\\[6pt]
1-\left(1-p(q)\right)^b, & q \notin c.
\end{cases}
\label{eq:exact_ratio_test}
\end{eqnarray}
This rule gives
\begin{eqnarray}
\mathrm{Pr}(c;M') = [\mathrm{Pr}(c;M)]^b,
\end{eqnarray}
and therefore exactly satisfies the ratio-test definition in Eq.~\eqref{eq:RT} of the main text.

\section{Gap test}
\label{app:GT}
Both gap test and ratio test compare the most and next most probable error configurations, but ratio test does so through the ratio of their log-likelihoods. Equivalently, ratio test may be viewed as a variant of gap test with an adaptive gap criterion. Specifically, the gap test can be recovered from the ratio test by replacing the fixed suppression factor $e^{-b}$ with the adaptive factor $[\,\mathrm{Pr}(c;M)]^{b-1}$.

For the practice gap test, with the error configuration $c$ from the first-round decoding, we modify the error model by taking 
\begin{eqnarray}
p'(q) =
\begin{cases}
\exp\left(-\frac{\ln p(q)}{\ln p(c)} b\right) p(q), & q \in c, 
\\[6pt]
p(q), & q \notin c,
\end{cases}
\label{eq:reweight_GT}
\end{eqnarray}
where $p(c) = \prod_{q \in c} p(q)$, and $b>0$ is the reweighing parameter.

We emphasize that in practice neither gap test nor ratio test provides an accurate detection of the true likelihood gap, because probabilities of error configurations other than $c$ are usually also changed.

\begin{figure}[h]
\centering
\includegraphics[width=1\linewidth]{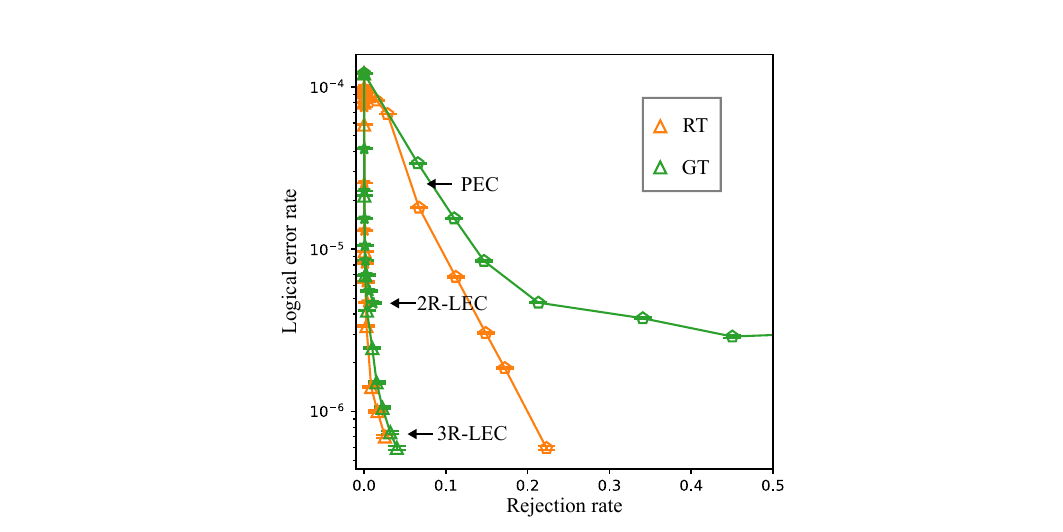}
\caption{
Comparison between the ratio test (RT) and gap test (GT). Performance is evaluated for the [[72,12,6]] BB code using the BP-OSD decoder. Each data point is estimated from $2.56 \times 10^9$ Monte Carlo shots under circuit-level noise ($p=0.001$) with $d=6$ parity-check measurement rounds. For RT, the reweighting parameter is set to $b = 1 + z$, with $z \in \{10^{-15}, 10^{-8}, 10^{-4}, 10^{-3}, 0.01, 0.1, 0.2, 0.3, 0.4, 0.5\}$; for GT, we take $b \in \{2, 5, 7, 10, 12, 15, 20, 25\}$.
}
\label{fig:LER_GT}
\end{figure}

The performance of the gap test is shown in Fig.~\ref{fig:LER_GT}. Following the definition in Eq.~\eqref{eq:reweight_GT}, we consider eight distinct values of the reweighting parameter: $b \in \{2, 5, 7, 10, 12, 15, 20, 25\}$. We apply PCHIP interpolation to the 3R-LEC data for the ratio test and gap test in Fig.~\ref{fig:LER_GT}, obtaining rejection rates of $r_R=5.48\times 10^{-4}$ and $r_G=6.72\times 10^{-4}$, respectively, for achieving a reduction in the logical error rate by one order of magnitude. This gives $r_R/r_G \approx 0.82$. Our results indicate that the ratio test slightly outperforms the gap test in terms of post-selection efficiency.

\section{Discrepancy between ideal reweighting and practical reweighting}
\label{app:IR_PR}
A discrepancy exists between ideal and practical reweighting. Specifically, for the ratio test, both approaches suppress the probability of the correction $c$ according to Eq.~\eqref{eq:RT} in the main text; however, they differ in their treatment of alternative error configurations consistent with the same syndrome. While ideal reweighting leaves the probabilities of these alternative error configurations unchanged, the practical operation [Eq.~\eqref{eq:reweight} in the main text] may alter them if they share elementary errors with $c$. Despite this deviation from the ideal case, practical reweighting remains effective because it consistently lifts the relative position of correction $c$ within the negative log-likelihood spectrum.

Using Tanner graph notation, we represent the check matrix as a tuple $(C, B, E)$, where $C$ is the set of nodes corresponding to parity checks, $B$ is the set of nodes representing elementary errors, and $E \subseteq C \times B$ specifies which elementary errors are detected by which checks. Each error configuration $e$ is a subset of elementary errors ($e \subseteq B$), and the correction is similarly a subset $c \subseteq B$. In the practical reweighting operation, we suppress the probabilities of the elementary errors in $c$, according to Eq.~\eqref{eq:reweight} in the main text. Then, for an arbitrary error configuration $e$, its negative log-likelihood under the modified model $M'$ is given by
\begin{equation}
-\ln\mathrm{Pr}(e;M') = -\ln\mathrm{Pr}(e;M) - (b-1)\sum_{q\in e\cap c}\ln p(q),
\end{equation}
where $-\ln\mathrm{Pr}(e;M)$ is the negative log-likelihood under the original error model $M$. The relative gap in negative log-likelihood between $e$ and $c$ under the modified model can then be expressed as
\begin{align}
[-\ln\mathrm{Pr}(e;M')] - [-\ln\mathrm{Pr}(c;M')] = [-\ln\mathrm{Pr}(e;M)] - [-\ln\mathrm{Pr}(c;M)] + (b-1)\sum_{q\in e\setminus c}\ln p(q).
\end{align}
Since $\sum_{q\in e\setminus c}\ln p(q)$ is always negative, the negative log-likelihood gap between $e$ and $c$ narrows as $b$ increases, eventually leading to a level crossing. This crossing is precisely the desired effect for detecting the original gap, although achieving the crossing point may require a larger value of $b$ compared with ideal reweighting.

\begin{figure}[t]
\centering
\includegraphics[width=1\linewidth]{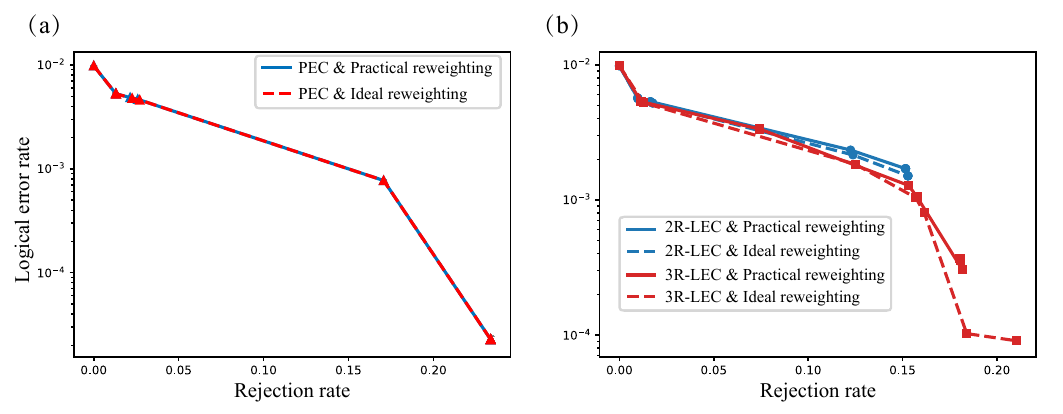}
\caption{
Discrepancy between ideal and practical reweighting. We simulate the [[7,1,3]] Steane code under a phenomenological noise model with $d=3$ rounds of parity-check measurements, where each bit is subject to independent bit-flip errors with probability $p=0.01$. Here, we using the maximum likelihood decoding. The reweighting parameter for ratio test is taken over the set $b \in \{1.2, 1.4, 1.6, 1.8, 2.0,2.2,2.4,2.6,2.8,3.0\}$. Each data point is estimated from $10^7$ Monte Carlo shots.
}
\label{fig:MLD_Ideal_vs_practical}
\end{figure}

To evaluate the quantitative deviation from ideal reweighting, we performed numerical simulations of the [[7,1,3]] Steane code. The results, presented in Fig.~\ref{fig:MLD_Ideal_vs_practical}, demonstrate that the ideal reweighting performs better and the discrepancy is small at low rejection rates.

\textit{Remark on normalization.} In the definitions of ideal and practical reweighting, we focus on errors compatible with the measured syndrome $s=S(c)$. This is sufficient for decoding, because errors with different syndromes are irrelevant to the decoding decision. Thus, in both ideal and practical reweighting, the probabilities of errors outside this syndrome class may be adjusted implicitly to ensure normalization of the full probability distribution.

\section{Shielding effect and the saturation phenomenon in LEC}
\label{app:explanation}

\subsection{Shielding effect}

The entire set of physical error configurations can be partitioned into equivalence classes according to their corresponding logical effects. Ideally, decoding is performed by evaluating the total likelihood of each class and selecting the most probable one. In practice, however, decoders typically identify the single most likely physical error (the correction) rather than the most likely class; this can be viewed as approximating the total likelihood of a class by the likelihood of the most probable physical error in the class. When the likelihood gap between the most-likely and next-most-likely classes is small, the correction is more prone to resulting in a logical error. Therefore, the goal of post-selection is to reject instances characterized by a narrow gap between these two leading classes.

Argument reweighting probes this likelihood gap by modifying the error model. A plausible explanation for the observed saturation phenomenon is a \textit{shielding effect}: high-likelihood physical errors belonging to the same logical class as the initial correction can effectively ``shield'' errors in competing classes. Consequently, these alternative classes remain undetected during the reweighting process, even when the true gap is small.

\begin{figure*}[h]
\centering
\includegraphics[width=0.8\linewidth]{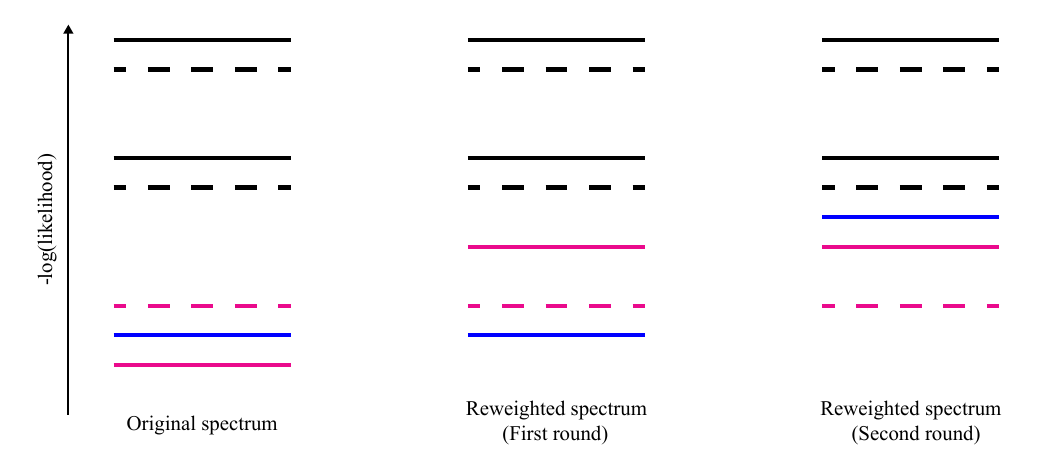}
\caption{
Negative log-likelihood spectrum of physical errors. Solid and dashed horizontal lines represent physical error configurations belonging to distinct logical equivalence classes.
}
\label{fig:spectrum}
\end{figure*}

This mechanism can be understood through the spectrum of negative log-likelihoods illustrated in Fig.~\ref{fig:spectrum}. As an example, consider a spectrum containing two logical classes, represented by solid and dashed lines, respectively. The most likely physical error (the red solid line) belongs to the ``solid'' class and is selected as the correction. Suppose that the next most likely physical error (the blue solid line) also belongs to the solid class, while the third most likely error (the red dashed line) belongs to the ``dashed'' class. If the red dashed line is close to the red solid line, the instance should ideally be rejected due to the small gap.

In the PEC scheme, reweighting lifts the red solid line (making it less likely). The decoder then selects the next most likely error, which is the blue solid line. Since the physical correction has changed, PEC correctly rejects the instance, though it does so based on the gap between physical errors rather than the gap between classes.

However, in 2R-LEC, the instance is accepted because the new most likely error (blue solid) still belongs to the same class as the original correction. Despite the narrow gap between the two classes (the proximity of the red dashed line to the red solid line), the instance remains accepted because the dashed class was shielded by the blue solid line. Due to this shielding effect, an instance may be accepted regardless of how high the red solid line is lifted. This explains the saturation phenomenon in LEC.

We note that argument reweighting, as implemented in practice [e.g., in Eq.~\eqref{eq:reweight} in main text], modifies the likelihood of all elementary errors within a correction. This collective modification of the error model may also contribute to the saturation phenomenon in addition to the shielding effect described above.

\subsection{Mitigation of the saturation phenomenon}

Shielding effect also explains the improvement offered by 3R-LEC. In the second round of reweighting, the blue solid line is also lifted. This finally allows the red dashed line from the alternative class to become the most likely correction. Consequently, 3R-LEC detects the competing class and rejects the instance.

\begin{figure*}[h]
\centering
\includegraphics[width=1\linewidth]{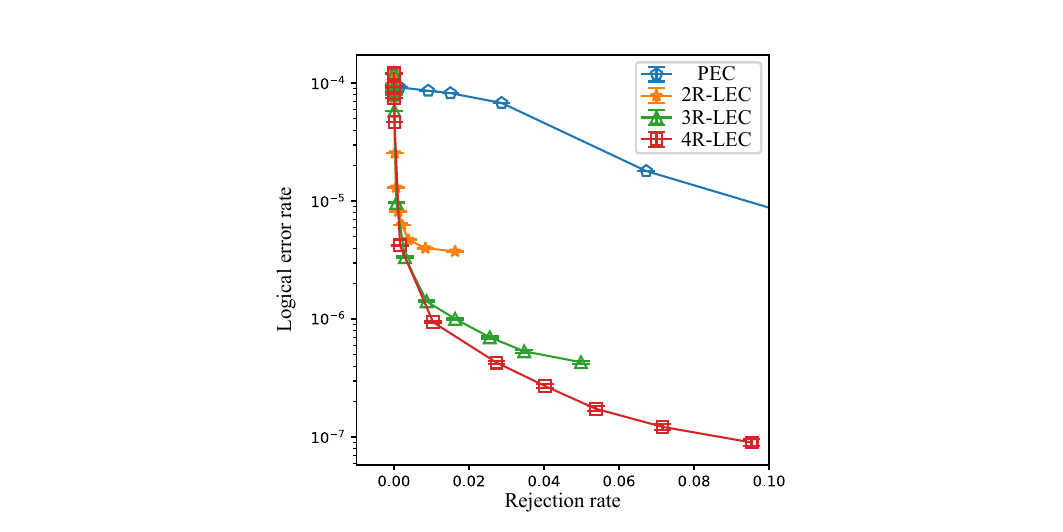}
\caption{
Performance with increasing reweighting-decoding rounds. Results are shown for the [[72,12,6]] BB code using the BP-OSD decoder. Each data point is estimated from $2.56 \times 10^9$ Monte Carlo shots under circuit-level noise ($p=0.001$) with $d=6$ parity-check measurement rounds. In post-selection, the reweighting parameter is set to $b = 1 + z$ for ratio test, with $z \in \{10^{-15}, 10^{-8}, 10^{-4}, 10^{-3}, 0.01, 0.1, 0.2, 0.3, 0.4, 0.5,0.7,1\}$.
}
\label{fig:LER_4RLEC}
\end{figure*}

While we focused on 2R-LEC and 3R-LEC in the main text, this reweighting-decoding procedure generalizes naturally to additional rounds. For example, by constructing a modified error model $M'''$ that suppresses the likelihoods of the preceding corrections $\{c, c', c''\}$, one can obtain a fourth-round output $c'''$. This defines the following enhanced decision rule:
\begin{itemize}
\item {\it Four-round logical-error criterion (4R-LEC)}---Accept iff $L(c''') = L(c'') = L(c') = L(c)$.
\end{itemize}
The performance of this 4R-LEC scheme is illustrated in Fig.~\ref{fig:LER_4RLEC}. As shown, 4R-LEC further improves post-selection efficiency and pushes the saturation threshold, albeit at the cost of increased runtime.

\begin{figure*}[h]
\centering
\includegraphics[width=1\linewidth]{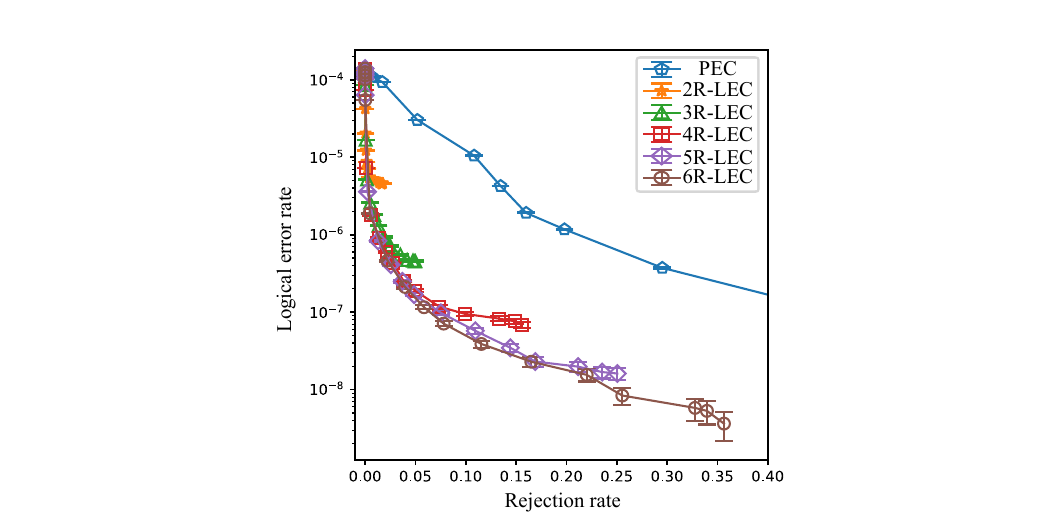}
\caption{
Performance with increasing reweighting-decoding rounds. Results are shown for the [[72,12,6]] BB code using the Relay-BP decoder. Each data point is estimated from $2.56 \times 10^9$ Monte Carlo shots under circuit-level noise ($p=0.001$) with $d=6$ parity-check measurement rounds. In post-selection, the reweighting parameter is set to $b = 1 + z$ for ratio test, with $z \in \{10^{-15}, 10^{-8}, 10^{-4}, 10^{-3}, 0.01, 0.1, 0.2, 0.3, 0.4, 0.5,0.7,1,1.5,2,3,4,5\}$.
}
\label{fig:LER_6RLEC}
\end{figure*}

\begin{figure*}[h]
\centering
\includegraphics[width=1\linewidth]{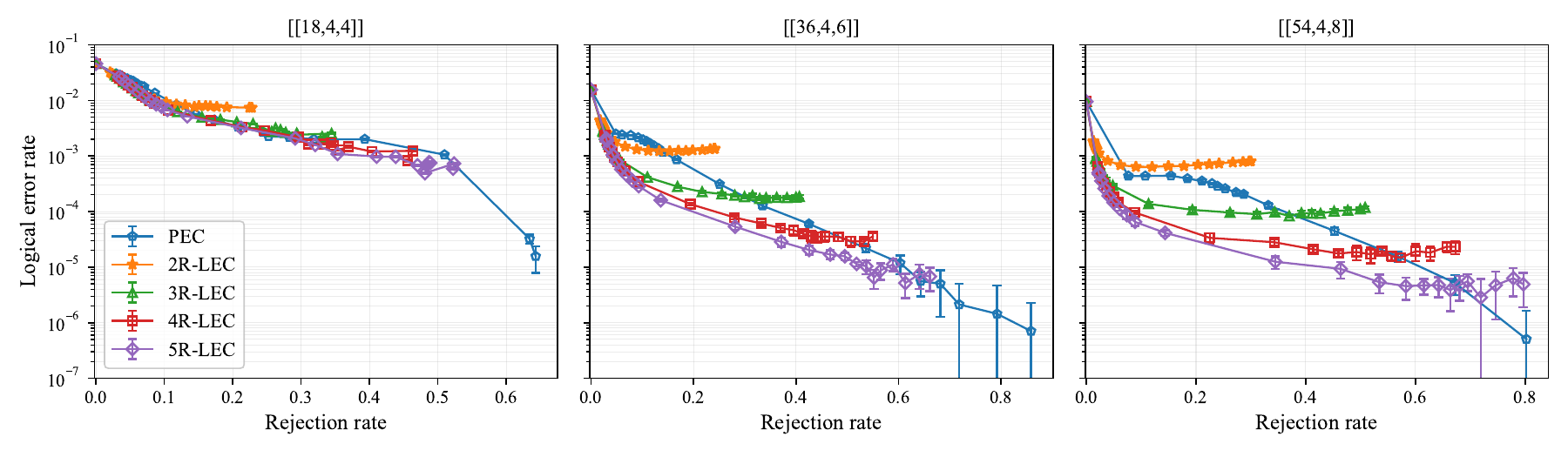}
\caption{
Performance with increasing reweighting-decoding rounds for BB codes of parameters [[18,4,4]], [[36,4,6]], and [[54,4,8]].
Here the BP-OSD decoder is used for all the codes. Each data point is estimated from $10^7$ Monte Carlo shots under circuit-level noise ($p=0.003$) with $d$ parity-check measurement rounds. In post-selection, the reweighting parameter is set to $b = 1 + z$ for ratio test, with $z \in \{10^{-12}, 10^{-6}, 10^{-3}, 10^{-3}, 0.01, 0.02, ..., 0.09, 0.1, 0.2, ..., 0.9, 1, 1.2, 1.5, 2, 2.5 \}$.
}
\label{fig:LER_5RLEC}
\end{figure*}

We can further increase the number of reweighting-decoding rounds to achieve continued reduction of the logical error rate. As shown in Fig.~\ref{fig:LER_6RLEC}, additional rounds provide significant gains up to 6R-LEC. Benchmarks across three different code instances confirm the robustness of this trend; see Fig.~\ref{fig:LER_5RLEC}. We note, however, that increasing the number of rounds does not reduce the logical error rate indefinitely, and the maximum suppression is ultimately bounded by the error rate associated with trivial syndromes.

\subsection{High-rejection-rate limit}

We now analyze the performance of reweighting in the high-rejection-rate limit. In the limit $b \to +\infty$, the likelihood of any non-empty correction $c$ vanishes. Consequently, a non-empty $c$ is never produced as the output of the second decoding round, i.e.,~$c \neq c'$. According to PEC, a circuit run is always rejected if $c \neq c'$. Therefore, in this limit, PEC only accepts runs where $c$ is empty, i.e.~no syndrome is observed; effectively, this doubles the code distance, representing the maximum possible suppression of the logical error rate~\cite{mitigating_smith_2024}.

\begin{figure}[h]
\centering
\includegraphics[width=1\linewidth]{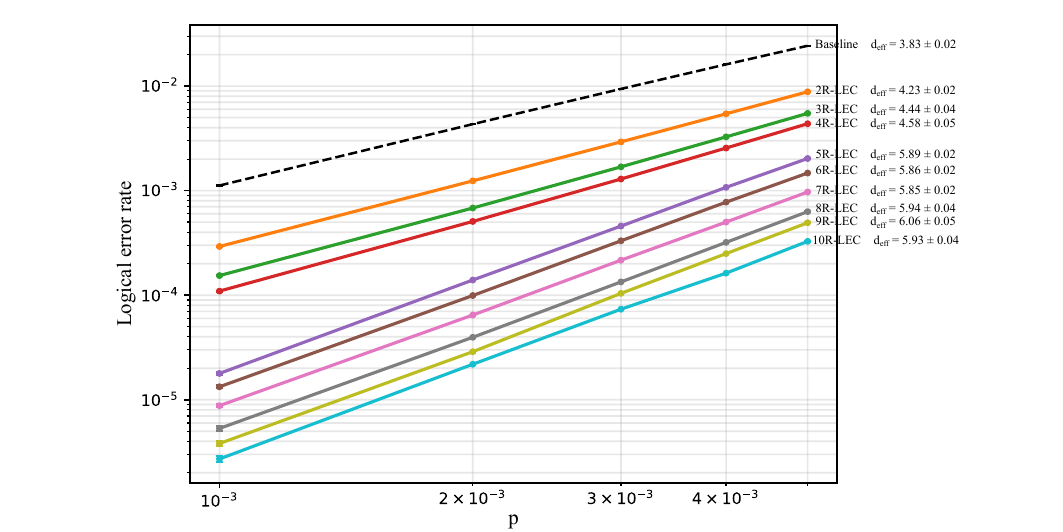}
\caption{
Effective distance in the LEC scheme with increasing decoding rounds. We evaluate the logical error rate of the $d=3$ rotated surface code decoded using the MWPM decoder integrated with up to nine rounds of reweighting-decoding (10R-LEC). Each data point is estimated from $10^8$ Monte Carlo shots under circuit-level noise ($p=0.001$) with $d=3$ parity-check measurement rounds. The reweighing parameter for ratio test is fixed at $b=2$ for all reweighting rounds. The effective code distance $d_{eff}$ is extracted by fitting the logical error rate to the scaling $P_L \propto p^{d_{eff}/2}$. The results demonstrate that while the 2R-LEC scheme is initially limited by the shielding effect, increasing the number of rounds allows the strategy to asymptotically approach the doubled-distance limit ($d_{eff} \to 2d = 6$).
}
\label{fig:LEC_limit}
\end{figure}

This argument does not apply directly to LEC, as runs with non-empty $c$ can still pass post-selection even as $b \to +\infty$ due to the shielding effect. Nevertheless, numerical results show that the LEC logical error rate decreases with the number of decoding rounds, eventually recovering the doubled-distance effect.

The shielding effect can be mitigated by increasing the number of reweighting-decoding rounds. Given a sufficiently large reweighing parameter $b$, each round ``lifts'' a specific physical error configuration in the high-likelihood region of the spectrum (see Fig.~\ref{fig:spectrum}). Provided that the number of rounds is sufficiently large, all high-likelihood physical errors belonging to the same logical class as the initial correction are eventually suppressed. This process allows the decoder to detect errors from alternative logical classes---unless the syndrome is trivial. Consequently, in the limit of large $b$ and numerous rounds, LEC recovers the distance-doubling effect of PEC. As illustrated in Fig.~\ref{fig:LEC_limit}, through repeated rounds of reweighting, the effective code distance $d_{eff}$, i.e.,~the slope in the log-log plot of the logical error rate versus the physical error rate, gradually approaches twice the original code distance ($2d$). Notably, the increased code distances arise from asymmetric rejection rates---higher at large $p$ and lower at small $p$. Adjusting $b$ to maintain a fixed rejection rate across all physical error rates may yield even larger effective code distances.

\section{Application to sliding-window decoding}
\label{app:SWD}
Sliding-window decoding~\cite{topological_dennis_2002,scalable_tan_2023,parallel_skoric_2023,lowlatency_gong_2024} is an important approach for real-time decoding, which is essential for the practical deployment of quantum error correction in quantum computation. In this scheme, syndrome data are decoded over successive finite time intervals, referred to as windows. Each window consists of $n_\mathrm{com}+n_\mathrm{buf}$ rounds of parity-check measurements and is divided into two parts: the first $n_\mathrm{com}$ rounds form the commit region, while the remaining $n_\mathrm{buf}$ rounds constitute the buffer region. The decoding algorithm is applied to the entire window, but the resulting correction is effectively committed only to the commit region; errors in the buffer region are deferred and corrected in the subsequent windows.

\begin{figure}[h]
\centering
\includegraphics[width=1\linewidth]{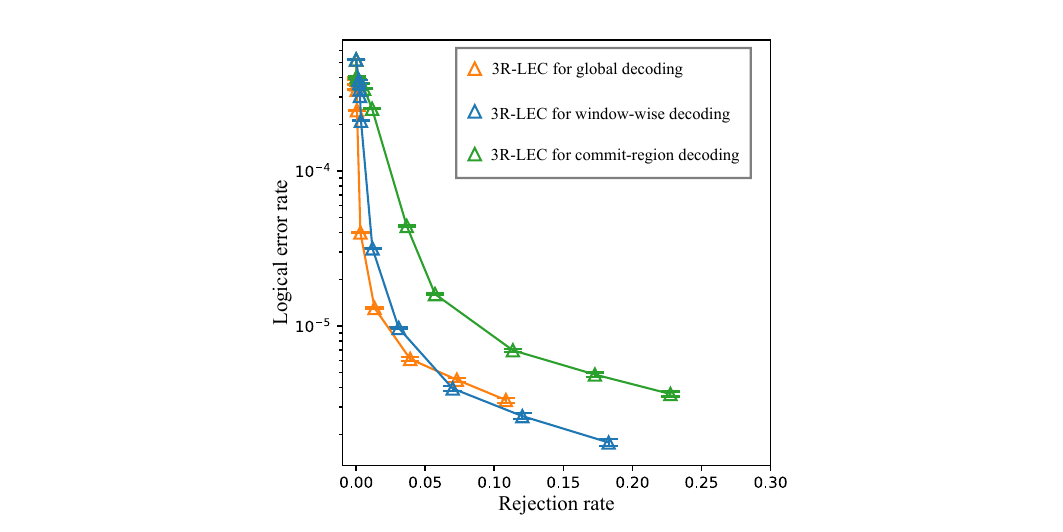}
\caption{
Performance of 3R-LEC under global and sliding-window decoding. Results show the logical error rate of the $[[72,12,6]]$ BB code after $4d$ rounds of parity-check measurements using the BP-OSD decoder, where $d$ is the code distance. For sliding-window decoding, each $2d$-round window is partitioned into a $d$-round commit region and a $d$-round buffer region. In post-selection, the reweighting parameter is set to $b = 1 + z$ for ratio test, with $z \in \{10^{-15}, 10^{-8}, 10^{-4}, 10^{-3}, 0.01, 0.1, 0.2, 0.3, 0.4, 0.5\}$. Each data point is estimated from $2.56\times10^8$ Monte Carlo shots, simulated under circuit-level noise ($p=0.001$).
}
\label{fig:LER_window}
\end{figure}

Argument reweighting seamlessly extends to sliding-window decoding. For each window, we reweight the error model across the entire window based on the correction from the previous decoding round of the same window. Post-selection is then performed using our proposed acceptance criteria. Numerical results for this sliding-window implementation are provided in Fig.~\ref{fig:LER_window}, demonstrating that argument reweighting in sliding-window decoding yields performance comparable to that in global decoding. 

Since each window comprises a commit region and a buffer region, our method can be applied flexibly to different regions. We evaluate two specific strategies: applying 3R-LEC to the decoded elementary errors across the entire window, or restricting the reweighting to those within the commit region. As illustrated in Fig.~\ref{fig:LER_window}, both approaches are effective, with the former being more efficient and achieving performance nearly identical to that of 3R-LEC under global decoding.

\section{Scalable application in fault-tolerant quantum computation}
\label{app:scalability}
The applicability of post-selection is limited by the rejection rate. Let $r$ denote the rejection rate per logical gate. A circuit with $N$ logical gates succeeds with probability $(1-r)^N \simeq e^{-Nr}$, implying $N \lesssim 1/r$ to maintain a reasonable overall success. Nevertheless, post-selection decoding can still be applied to large-scale fault-tolerant quantum computation through two approaches. First, it is well suited for small resource-state preparation and distillation circuits, such as those for magic states~\cite{Li_2015,magic_litinski_2019a,magic_gidney_2024,hiranoEfficientMagicState2025a,ghhp-cytl}. For large-scale circuits, the computation can be decomposed into smaller subcircuits implemented via gate teleportation~\cite{faulttolerant_gottesman_2014,timeefficient_yamasaki_2024,polylogtime_tamiya_2024,quantum_nguyen_2024,constantoverhead_zhang_2025a}, with post-selection applied to preparation circuits of gate-teleportation resource states; see Appendix~\ref{sec:gate_teleportation} for an example. Repeating each preparation independently allows the overall failure probability to be efficiently suppressed. Second, post-selection can be used for error mitigation on logical qubits~\cite{noiseagnostic_xie_2026a,error_zhou_2025,error_dinca_2025a}. In the spacetime noise inversion approach, ``super qubits''---logical qubits with error rates much lower than standard logical qubits---are required for error sampling, avoiding the complexity of characterizing the full error model~\cite{noiseagnostic_xie_2026a}. Because these sampling circuits are small and independent, post-selection can be applied to them efficiently to realize super qubits. The accuracy of the error-mitigated computation is then determined by the logical error rate of these super qubits, and post-selection effectively suppresses the computational bias to the level set by the logical error rate achieved under post-selection decoding.

\subsection{Gate teleportation}
\label{sec:gate_teleportation}

\begin{figure}[h]
\centering
\includegraphics[width=1\linewidth]{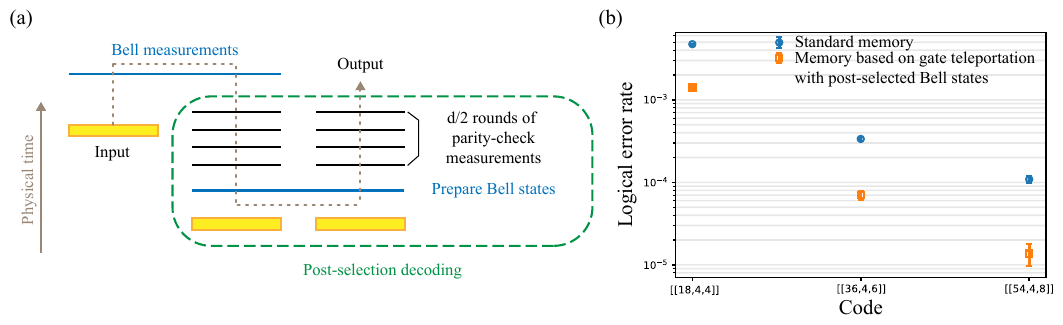}
\caption{
Parity-check measurements via gate teleportation with a post-selected Bell state.
(a) Protocol. Each yellow rectangle denotes a code block. By applying $d/2$ rounds of measurements to each auxiliary block of the Bell state, $d$ measurement rounds are effectively performed on the input block. The dotted cinereous line indicates the direction of effective time across the $d$ rounds.
(b) Performance comparison. Numerical results compare the teleportation protocol against a baseline of $d$ measurement rounds applied directly to the input block. Post-selection utilizes 3R-LEC with $b=1.5$ for ratio test. For the three codes, the corresponding rejection rates are $0.121$, $0.145$, and $0.197$, respectively. Each data point is estimated from $10^6$ Monte Carlo shots, simulated under circuit-level noise ($p=0.001$).
}
\label{fig:Gate_teleportationp}
\end{figure}

In fault-tolerant operations based on gate teleportation, the resource-state preparation circuit functions as an independent subroutine; consequently, post-selection is applicable to the preparation circuit. In Fig.~\ref{fig:Gate_teleportationp}, we present an example of this procedure and the resulting reduction in logical error rates.

In this example, we utilize gate teleportation to effectively implement $d$ rounds of parity-check measurements, which serve as the core component of logical operations such as code surgery. To achieve this, a logical Bell state is first prepared using two auxiliary blocks, each subject to $d/2$ rounds of parity-check measurements. Before using the resource state in gate teleportation, errors within these auxiliary blocks are processed via post-selection decoding (the first decoding phase); if an instance is rejected, the preparation circuit is re-executed. We then perform a transversal Bell measurement between one of these auxiliary blocks and the input block, while the other auxiliary block serves as the output. Syndrome information extracted from the transversal Bell measurement is subsequently used to correct errors on the input block via standard decoding without post-selection (the second decoding phase). Since $d/2$ rounds of measurements are performed on each auxiliary block, the total effective number of parity-check rounds applied to the input block is $d$.

The numerical results are presented in Fig.~\ref{fig:Gate_teleportationp}(b). We compare our post-selection protocol against a standard baseline where $d$ rounds of parity-check measurements are directly applied to the input block and processed via standard decoding. Our results show that the post-selection protocol suppresses the overall logical error rate to a level nearly equivalent to that of a single measurement round: the suppression factor is approximately $1/d$. Specifically, for the $[[54,4,8]]$ code, the logical error rate is reduced by nearly an order of magnitude.

The protocol presented here is practical for systems that support transversal CNOT gates (required for Bell state preparation and measurement) and possess long coherence times, such as trapped ions and neutral atoms. Note that the reduction in the logical error rate comes at the cost of increased runtime, as the input block must wait for the resource state to be successfully prepared. Suppose the entire preparation circuit requires a runtime $T$, the average waiting time is then $T/(1-r)$, where $r$ is the rejection rate. The largest rejection rate observed in numerical simulations is $0.197$ (see the caption of Fig.~\ref{fig:Gate_teleportationp}). Consequently, the time overhead---specifically the increase in overall circuit depth---due to post-selection is modest.

\section{Application to soft-information decoding}
\label{app:soft-info}
While we have focused on ``hard'' syndrome information---represented by binary vectors---argument reweighting is compatible with decoders that utilize soft syndrome information. This stems from the fact that our method operates solely on the decoder's error model for each shot. 

Decoding with soft information leverages the analog signals from measurement readouts to infer shot-specific noise strengths. In this regime, the error model is no longer static but becomes shot-dependent. Such readout signals contain richer information about the errors that occur in realistic devices; for instance, in superconducting systems, amplitude damping can lead to different readout error rates for physical qubits in the excited and ground states, resulting in different effective error models. Incorporating this soft information into the decoding process has been shown to improve the fault tolerance threshold under realistic noise conditions~\cite{improved_pattison_2021,soft_raveendran_2022}.

Our argument reweighting framework is fully applicable to these shot-dependent error models and integrates naturally with decoding schemes that utilize soft syndrome information. Let $M(s_\mathrm{soft})$ be the error model derived from the soft measurement outcomes $s_\mathrm{soft}$, and let $s$ denote the corresponding hard syndrome. A maximum-likelihood-type decoder identifies an initial correction $c = \text{Decoder}(s, M(s_\mathrm{soft}))$. The error model is then updated to $M'(s_\mathrm{soft})$ by suppressing the likelihood of the elementary errors constituting $c$; subsequently, the decoder is re-run to yield the second-round correction $c' = \text{Decoder}(s, M'(s_\mathrm{soft}))$. The remaining acceptance logic is identical to the hard-information case.

\section{Choice of $b$}

\begin{table}[t]
\begin{tabular}{ccccc}
\hline
$\eta$ & Code & BP-OSD & BP-LSD & Relay-BP  \\
\hline
\multirow{10}{*}{$10^{-1}$} & \multirow{3}{*}{[[72,12,6]]} 
& $b=1.1$ & $b=1.1$ & $b=1.2$  \\
\cline{3-5}
& & $6.816(5)\times10^{-4}$ & $1.5435(8)\times10^{-3}$ & $1.9681(9)\times10^{-3}$ \\
\cline{3-5}
& & $8(5)\times10^{-2}$ & $4.81(3)\times10^{-2}$ &  $3.71(3)\times10^{-2}$  \\
\cline{2-5}

& \multirow{3}{*}{[[90,8,10]]} 
& $b=1.0001$ & $b=1.001$ & $b=1.2$  \\
\cline{3-5}
&& $2.63(1)\times10^{-5}$ & $1.37(2)\times10^{-4}$ & $1.28(2)\times10^{-4}$ \\
\cline{3-5}
& & $9.9(2)\times10^{-2}$ & $9.11(8)\times10^{-2}$ & $9.1(3)\times10^{-2}$ \\
\cline{2-5}

& \multirow{3}{*}{[[144,12,12]]} 
& $b=1.0001$ & $b=1.001$ & $b=1.1$  \\
\cline{3-5}
&& $1.94(3)\times10^{-6}$ & $1.051(6)\times10^{-5}$ & $1.83(3)\times10^{-6}$ \\
\cline{3-5}
& & $9.5(7)\times10^{-2}$ & $9.1(3)\times10^{-2}$ & $7.3(12)\times10^{-2}$ \\
\hline

\multirow{10}{*}{$10^{-2}$} & \multirow{3}{*}{[[72,12,6]]} 
& $b=1.4$ & $b=1.4$ & $b=1.5$  \\
\cline{3-5}
&& $1.6315(3)\times10^{-2}$ & $2.6838(3)\times10^{-2}$ & $1.3314(2)\times10^{-2}$ \\
\cline{3-5}
& & $8.3(2)\times10^{-3}$ & $8.3(1)\times10^{-3}$ & $9.7(2)\times10^{-3}$
  \\
\cline{2-5}

& \multirow{3}{*}{[[90,8,10]]} 
& $b=1.2$ & $b=1.2$ & $-$  \\
\cline{3-5}
&& $7.154(5)\times10^{-4}$ & $2.802(1)\times10^{-3}$ & $-$  \\
\cline{3-5}
& & $5.3(4)\times10^{-3}$ & $7(2)\times10^{-3}$ & $-$ \\
\cline{2-5}

& \multirow{3}{*}{[[144,12,12]]} 
& $b=1.2$ & $b=1.2$ & $-$  \\
\cline{3-5}
&& $1.311(2)\times10^{-4}$ & $5.605(5)\times10^{-4}$ & $-$  \\
\cline{3-5}
& & $2.6(11)\times10^{-3}$ & $5.7(7)\times10^{-3}$ & $-$  \\
\hline
\end{tabular}
\caption{
The reweighing parameter $b$ required to reach target suppression factors $\eta$ for three BB codes under three different decoders. For each code, the top row shows the values of $b$ (taken from a desecrate set, see Ref.~\cite{note1} of the main text and Appendix~\ref{sec:parameters}) that achieve the target suppression, the middle row shows the corresponding rejection rates, and the bottom row lists the resulting suppression factors. These values are extracted from the same dataset used for Fig.~\ref{fig:Logical_error} in the main text.
}
\label{tab:3R-LEC}
\end{table}

The optimal choice of $b$ is fundamentally a trade-off determined by the requirements and hardware constraints: i) if the rejection rate is the primary bottleneck, $b$ should be set to the value that maximizes error suppression while remaining within a maximum permissible rejection budget; ii) if a low logical error rate is the primary goal and a high rejection rate is tolerable, $b$ should be chosen to meet the specific target logical error rate. As illustrated in Table~\ref{tab:3R-LEC}, the value of $b$ required to achieve a target suppression (e.g.,~a tenfold or hundredfold reduction in logical error rate) varies depending on both the specific code instance and the decoder employed.

In practice, $b$ can be determined through an offline calibration phase. For a given hardware platform, one can perform standard benchmarking tasks---such as randomized benchmarking or logical process tomography---to collect a set of error syndromes. The decoding can then be performed on this fixed experimental dataset for a sweep of $b$ values to map out the relationship between suppression and rejection. The $b$ value that satisfies the requirements is then selected for the actual computation. If the hardware noise is well-characterized, this calibration can also be performed via numerical simulations.

\begin{figure}[t]
\centering
\includegraphics[width=1\linewidth]{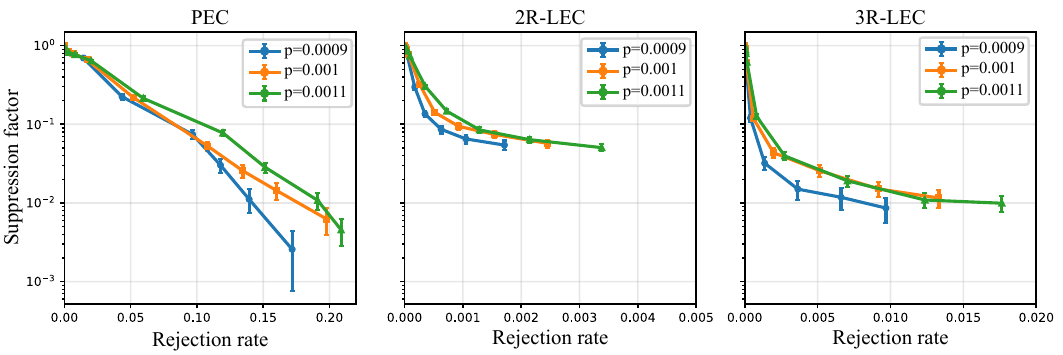}
\caption{
Performance of post-selection decoding under varying noise parameters. We evaluate the logical error suppression of the $[[72,12,6]]$ BB code subject to circuit-level noise with $10\%$ fluctuations around the physical error rate $p=0.001$. In post-selection, we set the reweighing parameter as $b = 1 + z$ for ratio test, with $z \in \{10^{-15}, 10^{-8}, 10^{-4}, 10^{-3}, 0.01, 0.1, 0.2, 0.3, 0.4, 0.5\}$. Each data point is estimated from $10^7$ Monte Carlo shots under circuit-level noise with $d=6$ parity-check measurement rounds.
}
\label{fig:Sensitivity}
\end{figure}

The above method for determining $b$ is robust if the underlying error model is stable. However, actual noise may vary with time. To investigate the sensitivity to the temporal drift of noise, we performed additional simulations to evaluate how performance fluctuates if the actual noise model deviates from the one used for calibration; see Fig.~\ref{fig:Sensitivity}. We find that the post-selection efficiency is relatively insensitive to minor variations in the noise model, suggesting that $b$ does not require high-frequency re-tuning.

\section{Additional numerical results}
\label{app:more_results}

\subsection{Comparison with the complementary gap method}
\label{app:CG}
\begin{figure}[h]
\centering
\includegraphics[width=1\linewidth]{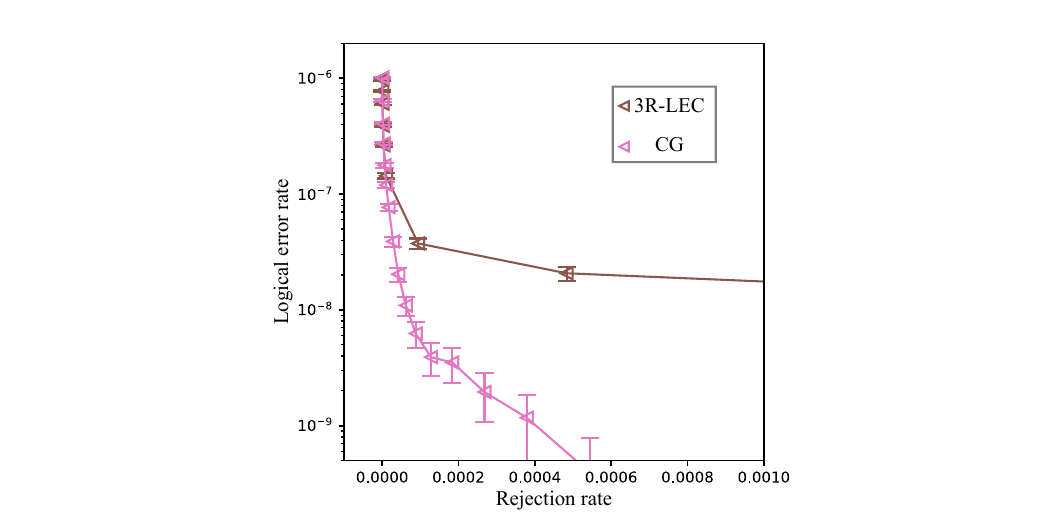}
\caption{
Comparison between 3R-LEC and the complementary gap (CG) method. Results are shown for the $d=9$ rotated surface code decoded using the MWPM decoder. Each data point is estimated from $2.56 \times 10^9$ Monte Carlo shots under circuit-level noise ($p=0.001$) with $d=9$ parity-check measurement rounds. For 3R-LEC, the reweighting parameter is set to $b = 1 + z$ for ratio test, with $z \in \{10^{-12}, 10^{-6}, 10^{-3}, 0.01, 0.02, 0.04, 0.06, 0.1, 0.2, 0.3, 0.4, 0.5, 0.6, 0.7, 0.8, 0.9, 1\}$; for CG, we sweep the threshold parameter from 1 to 100.
}
\label{fig:LER_CG}
\end{figure}

To compare 3R-LEC with the complementary gap method, we evaluate both strategies on a distance $d=9$ rotated surface code using the MWPM decoder. As illustrated in Fig.~\ref{fig:LER_CG}, 3R-LEC and the complementary gap method exhibit comparable efficiency in the low-rejection-rate regime. However, as the rejection rate increases, the complementary gap method begins to outperform 3R-LEC. We apply PCHIP interpolation to the data of 3R-LEC and the complementary gap method in Fig.~\ref{fig:LER_CG}, obtaining rejection rates of $r_{3R-LEC}=2.16\times 10^{-5}$ and $r_{CG}=1.28\times 10^{-5}$, respectively, for achieving a reduction in the logical error rate by one order of magnitude. This gives $r_{CG}/r_{3R-LEC} \approx 0.59$.

We have further performed numerical analysis to understand the difference between two methods, as shown in Fig.~\ref{fig:cg_ar_comparison}. As the value of complementary gap increases, the conditional logical error rate drops exponentially, which allows efficient rejection scheme. When we employ the 3R-LEC, we find that errorneous shots with small complementary gaps are also accepted. This provides a numerical support for our discussion in Appendix~\ref{app:explanation} that performance of LEC is currently limited by the shielding effect.

\begin{figure}[h]
\centering
\includegraphics[width=1\linewidth]{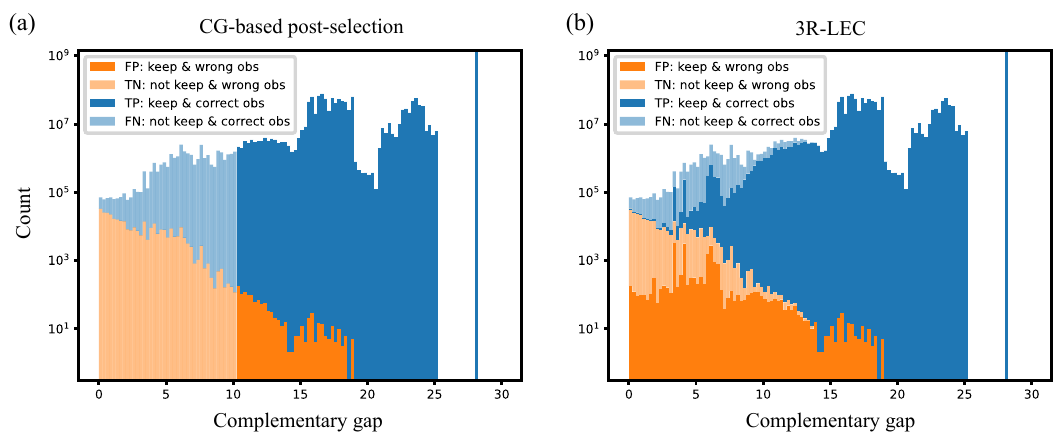}
\caption{
Histogram of success/failure by post-selection. We simulate $d$ rounds of parity-check measurements for a $d=5$ rotated surface code under circuit-level noise ($p=0.001$) using $2.56 \times 10^9$ Monte Carlo shots, with errors corrected via the MWPM decoder. Blue and orange colors indicate whether the prediction by decoder is correct or not. Thick and faint colors discriminate whether the shot was kept or not.
(a) Post-selection using the complementary gap method with the threshold parameter taken as 10.
(b) 3R-LEC using ratio test with reweighting parameter $b=2.5$.
The rejection rates are 0.012 and 0.014, respectively.
}
\label{fig:cg_ar_comparison}
\end{figure}

\subsection{Effect of reweighting on BP convergence}
\label{app:convergence}

\begin{figure}[t]
\centering
\includegraphics[width=1\linewidth]{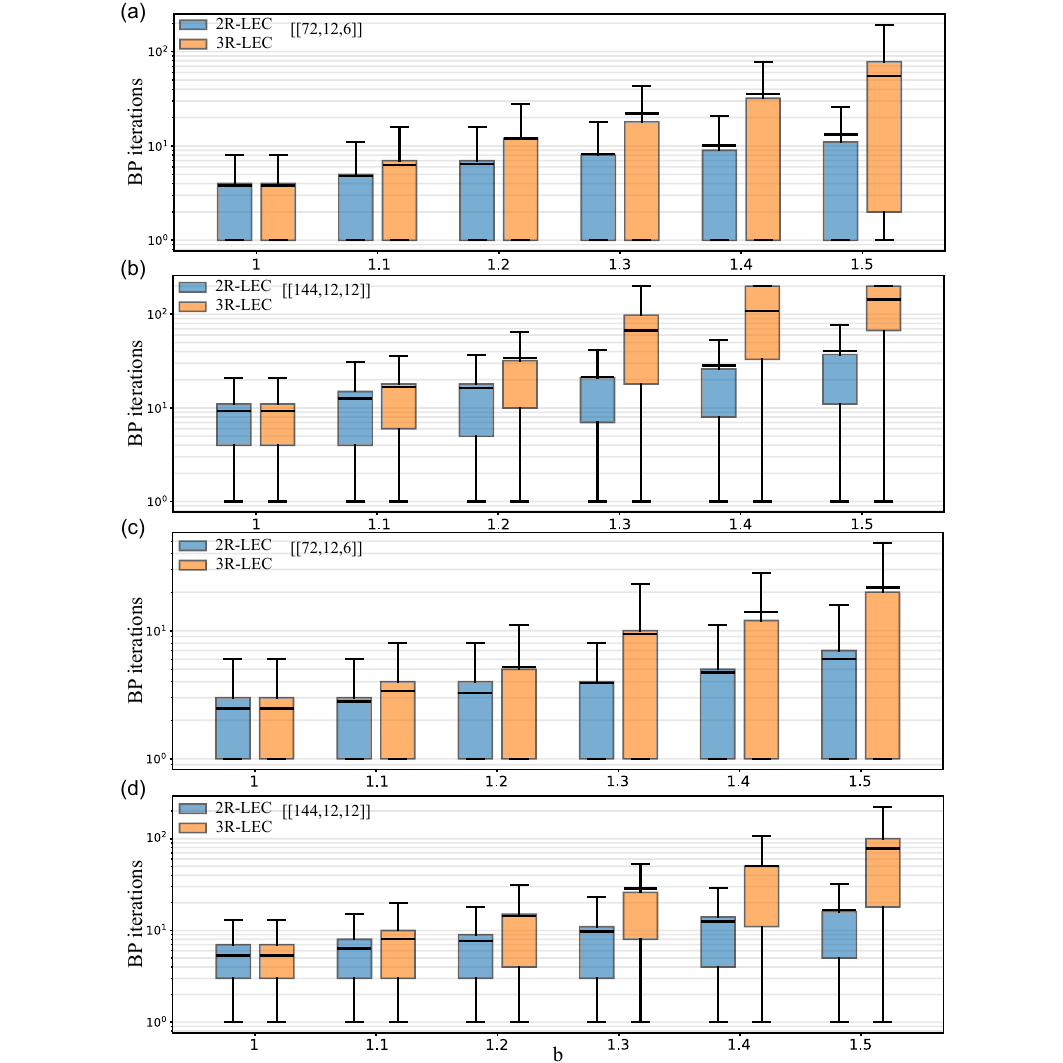}
\caption{
Effect of reweighting on BP convergence. 
(a), (b) Box and whisker plots of the number of BP iterations for the BP-OSD decoder under different reweighting parameters $b$ for ratio test. The maximum number of iterations is set to 200.
(c), (d) Box and whisker plots of the number of BP iterations for the Relay-BP decoder under different reweighting parameters $b$ for ratio test. The box and whisker plots follow a modified Tukey style, where the median line is replaced by the mean line. Each box and whisker plot is based on $10^6$ Monte Carlo shots, simulated under circuit-level noise ($p=0.001$) with $d$ parity-check measurement rounds.
}
\label{fig:iteration_shift}
\end{figure}

Reweighting the error probabilities perturbs the message-passing dynamics, generally leading to slower convergence. To investigate this effect, we performed numerical simulations of the $[[72,12,6]]$ and $[[144,12,12]]$ codes using both BP-OSD and Relay-BP decoders. The results are presented in Fig.~\ref{fig:iteration_shift}. In both decoders, we observe a shift in the iteration distribution toward higher values as the reweighting parameter $b$ increases in the ratio test configuration. Although the iteration count increases, the impact remains manageable: as shown in Table~\ref{tab:3R-LEC}, achieving a tenfold reduction in the logical error rate requires a reweighting parameter no greater than $b = 1.2$, which increases the mean iteration number by a factor of at most $3.69$ relative to the baseline decoder (corresponding to $b = 1$).

\subsection{Number of measurement rounds}

\begin{figure}[h]
\centering
\includegraphics[width=1\linewidth]{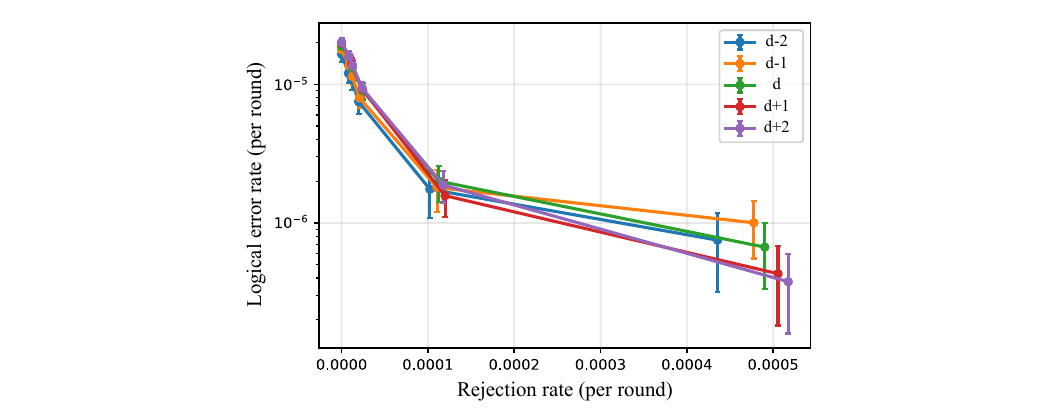}
\caption{
Logical error rate per parity-check measurement round. We simulate the $[[72,12,6]]$ BB code under circuit-level noise ($p=0.001$) using the BP-OSD decoder, with the number of measurement rounds ranging from 4 to 8. For these simulations, we calculate the rejection rate per round and the corresponding logical error rate per round. In post-selection, we take the reweighting parameter as $b=1+z$ for ratio test, where $z \in \{10^{-8}, 10^{-4}, 0.01, 0.1, 0.2\}$. Each data point is estimated from $10^6$ Monte Carlo shots.
}
\label{fig:different_cycle}
\end{figure}

In the main text, we report results for error-correction circuits consisting of $d$ rounds of parity-check measurements, where $d$ denotes the code distance. However, the efficacy of our post-selection approach does not depend on a specific number of measurement rounds. To demonstrate this robustness, we performed additional simulations for the $[[72, 12, 6]]$ BB code, varying the number of measurement rounds from 4 to 8. In Fig.~\ref{fig:different_cycle}, we plot the logical error rate per round versus the rejection rate per round for these cases. Aside from statistical fluctuations, we observe highly consistent trends in the suppression of the logical error rate across the different round numbers. These results indicate that the impact of the number of measurement rounds is insignificant.

\subsection{Performance at high physical error rate}

\begin{figure}[h]
\centering
\includegraphics[width=1\linewidth]{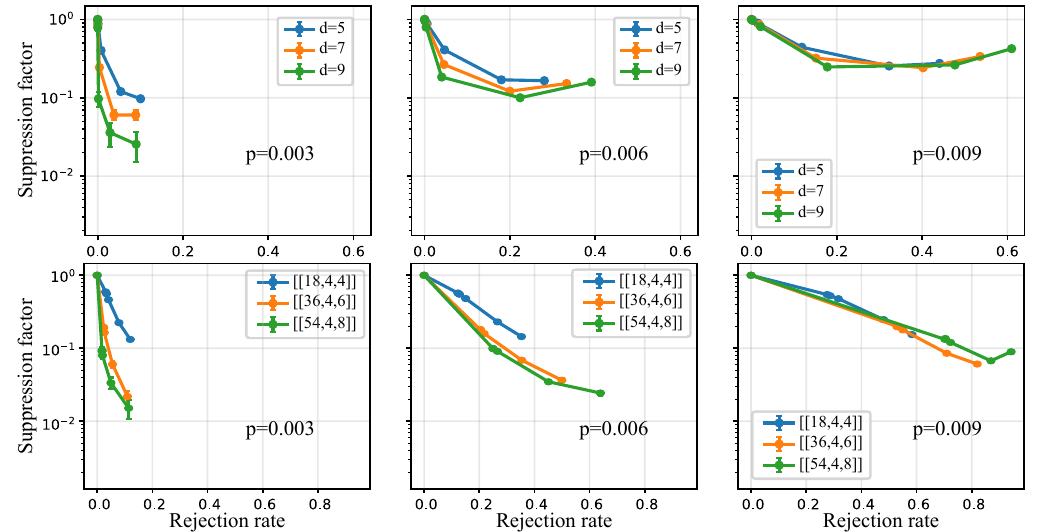}
\caption{
Performance of 3R-LEC at high physical error rates. We simulate circuit-level noise at three values higher than $p=0.001$, namely $p=0.003$, $0.006$, and $0.009$, and evaluate the resulting logical error suppression achieved by 3R-LEC for rotated surface codes and BB codes across different code distances. For decoding, we employ MWPM for the rotated surface codes and BP-OSD for the BB codes. In post-selection, the reweighting parameter is set to $b=1+z$ for ratio test, with $z \in \{10^{-3}, 0.01, 0.1, 0.3, 0.5\}$ for rotated surface codes and $z \in \{10^{-8}, 10^{-4}, 0.01, 0.1, 0.2\}$ for BB codes.
}
\label{fig:high_p}
\end{figure}

As the physical error rate $p$ increases, argument reweighting remains effective, but its efficiency decreases with increasing $p$. We benchmarked the 3R-LEC scheme for rotated surface codes and BB codes at $p=0.003$, $0.006$, and $0.009$. The results are illustrated in Fig.~\ref{fig:high_p}, which demonstrate that for a fixed code, achieving a target level of noise suppression requires a progressively higher rejection rate as $p$ increases.

Furthermore, these results confirm the favorable scaling reported in the main text (at $p=0.001$), where increasing the code distance $d$ enhances the efficiency of 3R-LEC. While this scaling behavior breaks down at the high physical error rate of $0.009$, the favorable trend remains robust across a wide range of noise levels.

\end{widetext}

\bibliography{references}

\end{document}